\RequirePackage{fix-cm}
\documentclass[smallextended]{svjour3}       
\smartqed  
\usepackage{graphicx}
\usepackage{bm}
\usepackage{amsmath}
\usepackage{amssymb}
\usepackage[countmax]{subfloat}
\usepackage{threeparttable}
\usepackage{enumitem}
\usepackage{color}
\usepackage{ulem} 
\begin{document}

\title{A non-perturbative study of the evolution of cosmic magnetised sources}



\author{I. Delgado Gaspar \and A. P\'erez Mart\'inez \and G. Piccinelli \and Roberto A. Sussman}


\institute{I. Delgado Gaspar \at
              Instituto de Investigaci\'on en Ciencias B\'asicas y Aplicadas, Universidad Aut\'onoma del Estado de Morelos, Av. Universidad 1001, Col. Chamilpa, 62210 Cuernavaca, MOR., M\'exico. \\
              \email{ismael.delgadog@uaem.edu.mx }           
           \and
           A. P\'erez Mart\'inez \at
              Instituto de Cibern\'etica Matem\'aticas y F\'isica (ICIMAF),  Calle E esq 15 No. 309 Vedado, La Habana 10400, Cuba.
              \email{aurora@icimaf.cu}
              \and G. Piccinelli \at
              Centro Tecnol\'ogico FES Arag\'on, Universidad Nacional Aut\'onoma de M\'exico, Avenida Rancho Seco S/N, 
              Bosques de Arag\'on, Nezaualc\'oyolt, Estado de M\'exico 57130, M\'exico.
              \email{itzamna@unam.mx}
              \and Roberto A. Sussman \at
              Instituto de Ciencias Nucleares, Universidad Nacional Aut\'onoma de M\'exico (ICN-UNAM), A. P. 70-543, 04510 M\'exico D. F., M\'exico.
              \email{sussman@nucleares.unam.mx}
}

\date{Received: date / Accepted: date}

\maketitle

\begin{abstract}
We undertake a hydrodynamical study of a mixture of tightly coupled primordial radiation, neutrinos, baryons, electrons and positrons, together with a gas of already decoupled dark matter WIMPS and an already existing ``frozen'' magnetic field in the infinite conductivity regime. Considering this cosmic fluid as the source of a homogeneous but anisotropic Bianchi I model, we describe its interaction with the magnetic  field by means of suitable equations of state that are appropriate for the particle species of the mixture between the end of the leptonic era and the beginning of the radiation-dominated epoch. Fulfilment of observational bounds on the magnetic field intensity yields a ``near FLRW'' (but strictly non--perturbative) evolution of the geometric, kinematic and thermodynamical variables. This evolution is roughly comparable to the weak field approximation in linear perturbations on a spatially flat FLRW background of sources in which the frozen magnetic fields are coherent over very large supra--horizon scales. Our approach and results may provide interesting guidelines in potential situations in which non--perturbative methods  are required to study the interaction between magnetic fields and the cosmic fluid. 
\end{abstract}

\section{Introduction}
A full account of the origin and evolution of cosmic and astrophysical magnetic fields is a long standing problem in contemporary theoretical physics. Magnetic fields have been observed in all scales: our own galaxy  \cite{MilkyWay}, low \cite{lowz} and high redshift \cite{highz} galaxies, and up to scales of galaxy clusters \cite{clusters} and superclusters \cite{superclusters} (for comprehensive reviews see \cite{OBS_review}). There is also indirect evidence, from gamma-ray observations of blazars \cite{Blazars} of coherent intergalactic magnetic fields in low density regions. While magnetic fields can result as a consequence of a variety of local astrophysical processes involving electric charges in motion 
(accretion into active galactic nuclei
or
compact objects, ionization of intergalactic gas and interactions between baryons and Cosmic Microwave Background (CMB) photons), all this multi--scale observational evidence supports the hypothesis that (at least) some of them may have a primordial origin from ``seed'' magnetic fields that emerged during some early evolution stage (or stages)  of the cosmic fluid. As shown extensively in recent comprehensive topic reviews \cite{Dario_review,Kandus:2010nw,Widrow:2012:rev,Durrer:2013:rev} on cosmological magneto--genesis, seed magnetic fields can be generated:
\begin{itemize}
\item  during inflation: by breaking the conformal invariance of Maxwell theory through the interaction of electromagnetic fields with dilaton--like scalar fields or axion-like pseudoscalar field \cite{non-conformal}, or by assuming conformal invariance with an FLRW background with negative spatial curvature \cite{negK}; 

\item in the various phase transitions after inflation by non--standard couplings that may generate helical magnetic fields on small scales (of the order of the horizon scale at these transitions). These fields can be amplified to all scales through an inverse cascade from the conservation of magnetic helicity \cite{invcasc}. Magnetic fields can also emerge through plasma mechanisms (Weibel effect) during QCD phase transition \cite{weibel}; 

\item during the classical plasma phase before recombination \cite{befrec}, by assuming that the photon--electron coupling in Compton scattering is much stronger than the coupling between photons and ions, leading to a density fluctuation field generating magnetic fields through pressure anisotropy and induced electric currents associated with the differences in the bulk electron and proton velocities.    
\end{itemize}
In the above mentioned articles on early times cosmological magneto--genesis the dynamics of the magnetic interaction 
in the cosmic fluid is examined  by means of a variety of techniques and theoretical frameworks depending on the specific transition: numerical simulations, 
Kinetic Theory, Statistical Mechanics, or gauge invariant vector perturbations over a spatially flat FLRW background (see details in 
\cite{Dario_review,Kandus:2010nw,Widrow:2012:rev,Durrer:2013:rev}, see other approaches to primordial magnetism in \cite{PRD75043004 (2007),PRD82.123007.2010,Piccinelli:2014dya}). However, a classical Maxwellian electromagnetic 
interaction (not necessarily restricted to early cosmic times) can also be incorporated into the full dynamics of General Relativity through 
an elegant first order system based on covariant objects defined in a 4--velocity frame (the 1+3 formalism \cite{Kandus:2010nw,Tsagas:2007yx,Tsagas}). In particular, this formalism is useful to study (following a perturbative or non--perturbative approach) the dynamics of a 
``frozen'' magnetic field without electric currents through the limit of infinite conductivity 
\cite{Tsagas:2007yx,Tsagas,Tsagas:1999ft,Tsagas:1999tu,Clarkson:2002dd,Tsagas2}. 
%
Whenever a spatially flat FLRW background is assumed in a perturbative treatment, a  ``weak field approximation'' can be defined in which the anisotropic stresses of the the magnetic field (associated with vector perturbations) are neglected and the latter field is described (at first order) as a sort of scalar source that dilutes as $B\sim a^{-2}$ \cite{Tsagas:2007yx,Tsagas,Tsagas:1999ft,Tsagas:1999tu,Clarkson:2002dd,Tsagas2,weakfield}. 
While this approximation is not gauge invariant, it leads to the same results (see proof in \cite{weakfield}) as gauge invariant perturbations when infinite conductivity is assumed. This weak field approximation was used also in earlier literature  \cite{Dario_review,Dario95,Dario96:1996} (see update in \cite{Kawasaki12,hortua2013}) in a purely FLRW context to examine the magnetic interaction at early cosmic times.

A different approach to cosmological magnetism that is non--perturbative (but restrictive) consists in considering a fluid (or fluid mixture) together with a magnetic field (in the infinite conductivity approximation) as sources of exact solutions of Einstein's equations. Work along these lines is found in older literature (see pioneering work in \cite{Zeldovich65}) considering very simple highly idealised fluids (even dust) in solutions with cylindrical symmetry \cite{Thorne67}, or homogeneous but anisotropic Bianchi models \cite{bianchis} and, in particular, spatially flat Bianchi I models  \cite{leblanc,Jacobs69,KC07}, which provide the simplest spacetime geometry compatible with the anisotropic stresses that characterise the magnetic interaction. More recently \cite{freestream}, a Bianchi I model was used to examine how the anisotropic stresses produced by free streaming neutrinos (described in terms of relativistic Kinetic Theory) allow for the fulfilment of the late time constraints placed by the CMB on the anisotropy of large scale (supra--horizon) magnetic fields\footnote{ We remark that Bianchi I models have received renewed attention \cite{recentB}, since present observational bounds from Planck and WMAP cannot rule out (in principle) the possibility that the large scale structure of the Universe is not perfectly isotropic from a statistical point of view.}.

In the present article we examine a magnetised early Universe cosmic fluid mixture as a source of a Bianchi I model. Instead of considering idealised fluids whose interaction with the magnetic field is unspecified (as in \cite{Thorne67,bianchis,leblanc,Jacobs69,KC07}), or a collision--less kinetic theory approach (suited to free streaming collision--less conditions as in \cite{freestream}), we consider fluid sources that satisfy physically motivated equations of state that are well suited for an early Universe cosmic fluid interacting with a magnetic field within a hydrodynamical regime, namely: a mixture of ideal gasses of magnetised fermions \cite{Chaichian:1999gd,Martinez:2003dz,Ferrer:2010wz} whose equation of state  reflects the full anisotropic effects of the magnetic field.  Evidently, this field  introduces anisotropic momentum fluxes that modify the energy--momentum tensor and thus affect the evolution of the state variables. Hence, the anisotropic pressure terms associated with the magnetic field necessarily contain a (``pure'') classical Maxwell term  \cite{Landau-Lifshitz}, but must also modify the equations of state of the fluid sources. We remark that in previous work we have studied the gravitational collapse of such magnetised sources in a Bianchi I geometry,  considering the case of zero  \cite{Alain_e-,Alain_2,Alain_n} and finite temperatures  \cite{Gaspar:2013eca}.

In order to examine the hydrodynamical evolution of the magnetised fluid mixture described above,  
we assume the magnetic field to be the only source of anisotropy of the energy--momentum tensor and consider two separate cosmic epochs where the effect of the magnetic field can be particularly relevant for a Universe full of free electric charges: (i) the end of the leptonic era, and: (ii) the beginning of the radiation dominated era immediately before cosmic nucleosynthesis.
Although all fermions, neutral and charged, interact with the magnetic field, we will assume an equation of state  in which the contribution of electrons and positrons is dominant. This is a good approximation, since in the cosmic times we are interested the constraint $eB \ll m^2_{p,n}$ holds and the interaction of the magnetic field with protons and neutrons can be neglected from the statistical properties of these particles. 

Besides introducing anisotropic stresses, primordial magnetic fields lead to various important effects on the evolution of early Universe sources. Hence, stringent bounds need to be imposed on their field strength in order to comply with numerous observational constraints, including limits of the abundances of small anisotropic stresses from older literature based on the COBE four-year anisotropy data \cite{Barrow:1997sy,Barrow:1997mj}. In particular, we are concerned on the effects of the magnetic field on cosmic nucleosynthesis, since it is the cosmological event in the cosmic stages under examination that is ``closest'' to us. 
Magnetic fields lead to various different effects on cosmic nucleosynthesis: (i) their contribution to the energy density content of the cosmic fluid affects the expansion rate; (ii) the electron-positron quantum statistics is modified, and so is the rate of neutron beta decay (see \cite{Dario_review} for a review). Considering these effects together, an upper bound of $ \left< B_0 \right> \leqslant 3 \times 10^{-7} \hbox{G}$ is necessary at length scales of the order of the Hubble horizon size at BBN time \cite{Dario95,Dario96:1996}. An updated value of this bound is: $\left< B_0 \right> \leqslant 1.5 \times 10^{-6} \hbox{G}$ (this is related to the local field amplitude $B$ contributed from all wavelengths)  \cite{Kawasaki12}. 
These bounds were obtained without considering the anisotropy introduced 
by the magnetic field. On the other hand, fitting CMB observations leads to a stronger constraint that appears to exclude
homogeneous cosmological magnetic fields much stronger than $10^{-9}$ G \cite{Kandus:2010nw,freestream}.

Besides considering observational bounds on the magnetic field, we also discuss how  a comparison of our approach and results can be made with those based on cosmological perturbations discussed and summarised in  \cite{Kandus:2010nw,Widrow:2012:rev,Durrer:2013:rev} and in previous work based on the weak field approximation \cite{Dario_review,Dario95,Dario96:1996,Kawasaki12,hortua2013}. This comparison is facilitated by our choice of a Bianchi I model that becomes a spatially flat FLRW model when the magnetic field vanishes ({\it i.e.} the magnetic field is the sole cause of anisotropy of the source), as in this case a sufficiently weak field can be regarded approximately as a sort of supra--horizon magnetic perturbation on a spatially flat FLRW background (see comments on this issue in \cite{freestream}), but it lends naturally with a comparison in the weak field regime studied in \cite{Tsagas2,weakfield,hortua2013} and in \cite{Dario_review}.  Without neglecting the importance of observational bounds, the obtained results may be useful in potential situations in which strong cosmic magnetic field need to be examined in a non--perturbative manner.  

The paper is organised as follows: in Sec. \ref{BI model} we examine the Einstein--Maxwell field equations that govern the dynamics of a magnetised cosmological fluid mixture in a Bianchi I geometry. The fluid sources and equations of state for the constituents of this magnetised fluid are discussed in detail in Sec. \ref{sources}. The Einstein--Maxwell system is transformed into a first order system given in terms of dimensionless variables in Sec. 4. The perfect fluid and FLRW limits are examined in Sec. 5, while in Sec. 6 we discuss the possible comparison with linear perturbations and the weak field regime and the late time evolution of the model. In Sec. \ref{results} we report our numerical results for the evolution of relevant thermodynamical and dynamical parameters. Finally, we provide our conclusions in Sec. \ref{conclusion}.

\section{Einstein--Maxwell equations for a Bianchi I model}\label{BI model}

Considering rectangular comoving coordinates $(t,x,y,z)$ with $x,\,y,\,z$ aligned along the principal axes of the shear tensor, Bianchi I models are described by the Kasner metric\footnote{Unless specified otherwise, we use natural units $G=c=1$.}:

\begin{equation}
ds^{2}=-dt^{2}+a_{1}^{2}\left(t\right)dx^{2}+a_{2}^{2}\left(t\right)dy^{2}+a_{3}^{2}\left(t\right)dz^{2}.\label{kasner}
\end{equation}
In order to describe a matter-energy source that includes magnetic interaction for this metric we consider the following energy--momentum tensor
\begin{equation}  
    T_{\mu\nu}=\rho \, u_\mu u_\nu + p\, h_{\mu\nu} 
    +\Pi_{\mu\nu} +\Lambda\, g_{\mu\nu},\label{Tenab}
\end{equation}
\noindent where $\rho=\rho(t),\,p=p(t)$ and $\Pi_{\mu\nu}$ are, respectively, the energy density, isotropic pressure and the traceless spacelike symmetric part of the stress tensor (a nonzero $\Pi_{\mu\nu}$ is necessary to describe a magnetic field, see equations (\ref{eq:Pzz}) and (\ref{eq:Pxx}) further ahead) and $\Lambda$ is the cosmological constant (whose value is subjected to observational constraints).  The Bianchi I geometry yields a uniform magnetic field (constant at each fixed $t$) of infinite extension along a unique spacelike tetrad vector direction that can always be identified with a space axis (say, the z-axis). Hence, in a comoving frame with $4$-velocity $u^{\mu}=\delta_{\,\,t}^{\mu}$ this stress-energy tensor takes the diagonal form:
\begin{equation}
T^{\mu}_{\,\,\nu}= \hbox{diag}\left[-\rho-\Lambda,P_{\bot}-\Lambda,P_{\bot}-\Lambda,P_{\|}-\Lambda\right],\label{eq:TabMatricial}
\end{equation}
where $P_{\bot}=P_{\bot}(t)$ and $P_{\|}=P_{\|}(t)$
are the pressures ({\it i.e.} eigenvalues of the tensor $\Pi_{\mu\nu}$) in the directions perpendicular and parallel to the magnetic field in the frame defined by the coordinates $(t,\,x,\,y,\,z)$.

The dynamics of a cosmological fluid allowing for a magnetic interaction follows from the coupled Einstein-Maxwell field equations ($G_{\mu\nu}=8\pi T_{\mu\nu}$), which for the Bianchi I geometry (\ref{kasner}) take the form:
\begin{eqnarray}
-G^{\,\,x}_{x}=\frac{\dot a_{2} \dot a_{3} }{a_{2}a_{3}}+\frac{\ddot a_{2} }{a_{2}}+\frac{\ddot a_{3}}{a_{3}}=-8\pi (P_{\bot}-\Lambda),\label{eq:Gxx}
\\
-G^{\,\,y}_{y}=\frac{\dot a_{1}\dot a_{3}}{a_{1}a_{3}}+\frac{\ddot a_{1}}{a_{1}}+\frac{\ddot a_{3}}{a_{3}}=-8\pi (P_{\bot}-\Lambda),\label{eq:Gyy}
\\
-G^{\,\,z}_{z}=\frac{\dot a_{1} \dot a_{2} }{a_{1}a_{2}}+\frac{\ddot a_{1}}{a_{1}}+\frac{\ddot a_{2}}{a_{2}}=-8\pi (P_{\|}-\Lambda),\label{eq:Gzz}
\\
-G^{\,\,t}_{t}=\frac{\dot a_{1} \dot a_{2} }{a_{1}a_{2}}+\frac{\dot a_{1} \dot a_{3} }{a_{1}a_{3}}+
\frac{\dot a_{2} \dot a_{3} }{a_{2}a_{3}}=8\pi (\rho+\Lambda).\label{eq:Gtt}
\end{eqnarray}
where $\dot{a}=u^{\alpha}a_{,\alpha}=a_{,t}$. The conservation equation $T^{\mu\nu}\,_{;\nu}=0$ yields:
\begin{equation}\label{EQU}
\dot{\rho}=-\left(\frac{\dot a_{1}}{a_{1}}+\frac{\dot a_{2}}{a_{2}}\right)\left(P_{\bot}+\rho \right)-
\frac{\dot a_{3}}{a_{3}}\left(P_{\|}+\rho \right),
\end{equation}
together with the only non--trivial of Maxwell's equations ($F^{\mu\nu}\,_{;\nu}=0$ and $F_{\left[\mu\nu;\alpha\right]}=0$):
\begin{equation} \label{MEq1}
\frac{\dot a_{1} }{a_{1}}+\frac{\dot a_{2}}{a_{2}}+\frac{\dot{B}}{B}=0.
\end{equation}
Notice that $B$ is not a scalar. It can be characterised as the single nonzero component of the covariant magnetic field vector in the comoving frame:  $B_\mu=B\delta_\mu^z=(1/2)\eta_{\mu\alpha\beta} F^{\alpha\beta}$, where $\delta_\mu^z$ and $\eta_{\mu\alpha\beta}$ are the Kronecker delta and Levi--Civita tensors. 
%

\section{Physically motivated field sources}\label{sources}

We will assume as the field sources of the Bianchi I model (i) a tightly coupled gas mixture of leptons, baryons and photons, (ii) an already decoupled non-interacting gas of cold dark matter WIMPS whose contribution is basically the rest--mass energy density (whose abundance value can be assumed to be very subdominant: $\Omega_{\textrm{\tiny{CDM}}}\sim10^{-6}$), (iii) an already existing magnetic field $B$. We will follow a purely phenomenological approach to this magnetic field, assuming that it emerged from previous fundamental processes whose details are beyond the scope of this paper (see brief description on cosmic magneto--genesis in the introduction and a comprehensive discussion  in \cite{Kandus:2010nw,Widrow:2012:rev,Durrer:2013:rev,Dario_review}) 
\footnote{Assuming a present day value $\Omega_0^\Lambda\sim 0.7$ for a $\Lambda$CDM background, the contribution of the cosmological constant is absolutely negligible in the cosmic times under consideration, hence we omit it until we discuss the late evolution of the models in Sec. \ref{latetimes})}.

We will also assume that the magnetic interaction is only significant with some of the particle species in (i). The total energy density and pressures for the source (\ref{Tenab})--(\ref{eq:TabMatricial}) have the form: 
\begin{eqnarray}
\rho=\rho^B + \rho^\gamma+\rho^\nu+\rho^{\textrm{\tiny{CDM}}}+\sum_{k} \rho^k,\label{rhoTot}
\\
P_{\bot}=P_{\bot}^{\,\,B}+P^{\,\,\gamma}+P^{\,\,\nu}+\sum_{k} P_{\bot}^{\,k},\label{PbotTot}
\\
 P_{\|}=P_{\|}^{\,\,B}+P^{\,\,\gamma}+P^{\,\,\nu}+\sum_{k} P_{\|}^{\,k},\label{PparTot}
\end{eqnarray}
where $k$ runs over all particles that could interact with the magnetic field, and the upper indices $B$, $\gamma$ and $\nu$ respectively denote the density and pressures associated with the magnetic field, the photons and the neutrinos.
%


 %
Because of the symmetries of the Bianchi I model the magnetic field in a comoving frame is necessarily homogeneous ({\it i.e.} purely time dependent) and (as we have chosen it) to be everywhere oriented along the $z$-direction. The magnetic density and pressures are then
\begin{equation}
P_{\bot}^B=-P_{\|}^B=B^2/(8\pi)=\rho^B,\label{PB} 
\end{equation}
leading to a positive pressure term along the $x$ and $y$-axes and a negative pressure along the field direction. This negative pressure can be interpreted as a tension or elasticity of the field lines, which tend to remain as ``straight'' as possible by reacting to any effect that distorts them \cite{Parker,Mestel}.
Consequently, an anisotropic pressure must give rise to an anisotropic expansion law.

Photons contribute to the source  with an isotropic pressure and energy density given by the equation of state
\begin{equation} p^\gamma(T) = \rho^\gamma(T)/3,\qquad  \rho^\gamma(T)=\emph{a}_{\mathcal{B}} T^4,\label{photEoS}\end{equation}
where $\emph{a}_{\mathcal{B}}=\pi^2/15$ and $T$ is the temperature. On the other hand the isotropic pressure and energy density of neutral leptons (neutrinos and antineutrinos) are given by \cite{Weinberg} 
\begin{equation} p^{\nu}(T) = \rho^{\nu} (T)/3,\qquad \rho^{\nu}(T) = 7 g^{\prime} \emph{a}_{\mathcal{B}} T^4 /16,\label{nuEoS}\end{equation}
with $g^{\prime}$ equals to $2\times3$, hence we have taken into account  the three different species of neutrinos and antineutrinos.

It is safe to neglect the interaction of photons, neutrinos and WIMP's with the magnetic field. However, leptons and baryons could interact with the field through their charges (if they are charged) and via their anomalous magnetic moment (if they are neutral). In any case the interactions with the field (via charges or anomalous magnetic moment)  lead to a momentum--energy tensor with anisotropic stresses \cite{Chaichian:1999gd,Ferrer:2010wz}.
We assume hereafter that the only particle species interacting with the magnetic field are electrons and positrons (charged leptons), protons (charged baryons) and neutrons (neutral baryons).

In general the  equation of state  for these species  in presence of a time--dependent magnetic field can be written as follows  \cite{Chaichian:1999gd,Ferrer:2010wz,Can1}:
%
\begin{eqnarray}
P_{\bot}^{k}&=& -\Omega^{k}-BM^{k},\label{eq:Pzz}
\\
P_{\|}^{k}&=& -\Omega^{k},\label{eq:Pxx}
\\
\rho^{k}&=& -\Omega^{k}+TS^k+\mu_k N^k.\label{eq:Uep}
\end{eqnarray}
%
where the upper index $k$ denotes generically the electron, positron, proton and neutron, $M^{k}=-\left(\partial\Omega^{k}/\partial B \right)$ is the  magnetisation,
$S^k=-\left(\partial\Omega^{k} / \partial T \right)$ is the entropy,
$N^k=-\left(\partial\Omega^{k} / \partial \mu_k \right)$ is the particle number density (with $\mu_k$ the chemical potential), and $\Omega^{k}$ the thermodynamical  potential which has two contributions:
\begin{equation}\label{contrib Omega}
 \Omega^{k}=\Omega^{k}_{\textrm{\tiny{SQFT}}}(B,\mu_k,T)+\Omega^{k}_{\textrm{\tiny{QFT}}}(B),
\end{equation}
where $\Omega^{k}_{\textrm{\tiny{SQFT}}}$ is the statistical Quantum Field Theory contribution and $\Omega^{k}_{\textrm{\tiny{QFT}}}$ (which does not depend on the temperature and the chemical potential) is the well-known Quantum Field Theory vacuum term given by
\footnote{See definitions in \cite{Ferrer:2010wz}. The term $\Omega^{k}_{\textrm{\tiny{QFT}}}$ has non-field-dependent ultraviolet divergencies, after renormalisation the Schwinger expression is obtained \cite{Schwinger}.}
\begin{equation}
\Omega^k_{\textrm{\tiny{QFT}}}(B)=-\frac{1}{4\pi^2}\sum\limits_{\eta=1,-1}\int\limits_{-\infty}^{\infty} dp_{\parallel}d^2p_{\perp}\varepsilon_k,\\\label{Omega Vn}
\end{equation}
and the statistical term:
\begin{eqnarray}
\Omega^{k}_{\textrm{\tiny{SQFT}}}(B,T,\mu)\hspace{-0.1cm} &=&\nonumber\\
&=&\hspace{-0.1cm}-\frac{1}{4\pi^2\beta}\hspace{-0.2cm}\sum\limits_{\eta=1,-1}\hspace{-0.1cm}\int\limits_{-\infty}^{\infty}
\hspace{-0.2cm}dp_{\parallel}d^2p_{\perp}\ln
\left(\left(1+e^{-\beta(\varepsilon_k-\mu)}\right)\left(1+e^{-\beta(\varepsilon_k+\mu)}\right)\right).\nonumber
\\
\label{Omega Statiticaln}
\end{eqnarray}
In the previous equations $\beta=1/T$, $\eta=1,-1$ correspond to the two orientations of the magnetic moment (parallel and antiparallel) with respect 
to the field, while $\varepsilon_k$ is the spectrum of the fermions given by:
\begin{equation} \label{fermionspectrum}
\varepsilon_k=\Bigg \{
\begin{array}{lr}
  \sqrt{p_{\parallel}^{2}+2|eB|l+m_k^{2}}   & \hbox{Charged fermions},
     \\
  \sqrt{p_{\parallel}^2+\left ( \sqrt{p_{\perp}^2+m_k^2} + \eta qB\right )^2}  &  \hbox{Neutral fermions}.
  \\
\end{array}
\end{equation}
In the equations above $m_k$ denotes the fermion mass and $q$ is the anomalous magnetic moment.
For charged fermions we need to carry on the following substitution:
\begin{equation}
 \int\frac{d^2p_{\perp}}{(2\pi)^2} \rightarrow  \frac{|eB|}{2\pi}\sum_{l=0}^{\infty}(2-\delta_{l0}),\label{subst}
\end{equation}
where $d(l)=2-\delta_{l0}$ is the spin degeneracy of Landau levels with $l\neq0$. For electron--positron pairs we assume a negligible chemical potential ($\mu=0$). Hence, the last term on the right-hand side of equation (\ref{eq:Uep}) vanishes.

For the temperature of the full duration of the stage of cosmic evolution we are interested the protons and neutrons satisfy  ($eB\sim T ^2 \ll m_{p,n}^2$), so they contribute to the evolution mostly through their rest energy. Therefore,  we assume that the only particles that contribute to the anisotropy in the   pressures are the electrons and positrons.

\section{Einstein--Maxwell equations as a first order system}

For a hydrodynamical numeric framework it is necessary to transform the second order Einstein--Maxwell equations (\ref{eq:Gxx})--(\ref{MEq1}) into a first order system of evolution equations for local kinematic covariant objects (see Appendix \ref{KVariables}) and the relevant state variables. We also need to introduce the following dimensionless evolution parameters: $\tau$,  ${\mathcal{H}}$, $S_{2}$ and  $S_{3}$
given by
\begin{eqnarray}
 \frac{d}{d\tau}&=&\frac{1}{H_*}\frac{d}{dt},\qquad \tau = H_*\,(t-t_i),\label{dimless1}\\
{\mathcal{H}}&=&\frac{\theta/3}{H_*},\quad  S_{2}=\frac{\Sigma_{2}}{H_*},\quad  S_{3}=\frac{\Sigma_{3}}{H_*},\label{dimless2}
\end{eqnarray}
where $H_*$ is an inverse length defined by the condition $3H_*^{2}=8\pi G\lambda/c^4$, where $\lambda= m_e/ \left(\pi^{2} \lambda_c^3\right)$ (with $\lambda_c$ the Compton wavelength and $m_e$ the electron mass),  $\theta=u^\mu\,_{;\mu}$ is the expansion scalar and $\Sigma_2,\,\Sigma_3$ are the eigenvalues of the shear tensor (see Appendix A).   Notice that $H_*$ roughly corresponds to the Hubble length in the outset of the radiation epoch (just before nucleosynthesis).

We also define for generic magnetised gas mixtures the following  dimensionless variables
\begin{eqnarray} \mathcal{E} = \frac{\rho}{\lambda},\qquad \mathcal{P}_\bot = \frac{P_\bot}{\lambda},\qquad \mathcal{P}_\| = \frac{P_\|}{\lambda},\label{LasGamas}\\
 \mathcal{B} =\frac{B}{B_{c}}, \quad  \mathcal{T}=\frac{T}{m_e},  \label{eq:func adimen}\end{eqnarray}
where $\rho,\,P_\bot,\,P_\|$ are the total energy density and anisotropic pressures in (\ref{rhoTot})--(\ref{PparTot}) and $B_c = m^{\,\,2}_e/e=4.41 \times 10^{13} G$ is the critical magnetic field for electrons\footnote{The critical magnetic field for an electron as defined above is the strength at which electron cyclotron energy equals its rest energy.}.
Considering the variables (\ref{dimless1})--(\ref{eq:func adimen}), the Einstein-Maxwell field equations become the following first order system:
%
\begin{eqnarray} \label{EDin}
\mathcal{H}_{\tau}&=&-\frac{1}{2}\left(3 \mathcal{E}+2\mathcal{P}_{\bot}+\mathcal{P}_{\|}\right)-\frac{1}{2}\left[\left(S_2+S_3\right)^2 -S_2 S_3\right],\label{EHt}
\\
S_{2,\tau}&=&\mathcal{P}_{\perp}-\mathcal{P}_{\|}-3\mathcal{H}S_{2},\label{EqS2}
\\ 
S_{3,\tau}&=&2\left(\mathcal{P}_{\|}-\mathcal{P}_{\perp}\right)-3\mathcal{H}S_{3},\label{EqS3}
\\
\mathcal{B}_{,\tau}&=& \mathcal{B} \left(S_{3}-2\mathcal{H}\right),\label{EqB}
\\ 
\mathcal{T}_{,\tau}&=&-\frac{1}{\mathcal{E}_{,\mathcal{T}}}\left(
\left(2\mathcal{H}-S_3\right)\left(\mathcal{E}+\mathcal{P}_{\perp}\right)+
\left( \mathcal{H}+S_3\right)\left(\mathcal{E}+\mathcal{P}_{\|}\right)-
\mathcal{E}_{,\mathcal{B}} \mathcal{B} \left(S_{3}-2\mathcal{H}\right)
\right),\nonumber
\\ \label{EqT}
\end{eqnarray}
together with the Hamiltonian constraint
\begin{equation}
3\mathcal{E} = -S_{2}^{2}-S_{3}^{2}-S_{2}S_{3}+3\mathcal{H}^{2}, \label{constraint}\end{equation}
where $\mathcal{E}_{,\mathcal{T}}=\partial\mathcal{E}/\partial\mathcal{T}$ and $\mathcal{E}_{,\mathcal{B}}=\partial\mathcal{E}/\partial \mathcal{B}$ in (\ref{EqT}). 

Since we have provided in (\ref{PB})--(\ref{subst}) the equations of state for all the constituents of the total energy density and pressures in (\ref{rhoTot})--(\ref{PparTot}), we can now provide constraints that link $\mathcal{E},\,\mathcal{B},\,\mathcal{P}_{||},\,\mathcal{P}_\perp$  (or $\rho,\,B,\, P_\|,\,P_\bot$), the differential equations (\ref{EHt})--(\ref{EqT}) and the constraint (\ref{constraint}) become a complete and self-consistent system whose numerical integration allows us to examine the dynamical evolution of a magnetised Universe in the cosmic stages we are concerned. Since all thermodynamical functions depend only on $\mathcal{B}$ and $\mathcal{T}$ (dimensionless magnetic field and temperature, respectively), from the numerical solutions for these two variables we can obtain the thermodynamical functions. On the other hand, the solutions for $S_2$, $S_3$ and $\mathcal{H}$ provide the necessary information to study the kinematical evolution of the cosmic fluid. The local proper volume can be expressed in terms of ${\mathcal H}$ as follows:
\begin{equation}\label{localVOL}
V(\tau)=V(\tau_i) \exp \left( 3 \int _{\tau_i}^{\tau}{\mathcal H} d\tau\right),
\end{equation}
and the expression for metric coefficients reads:
\begin{equation}\label{Qs}
a_{\rm{a}}(\tau)=a_{\rm{a}} (\tau_i) \exp \left( \int _{\tau_i}^{\tau}\left( S_{\rm{a}} + {\mathcal H}\right)d\tau\right), \quad \rm{a}=1,2,3
\end{equation}
with $S_1=-(S_2+S_3)$ (since the shear tensor is trace--free).

\section{The FLRW and perfect fluid limits}\label{LimitRW}
If the magnetic field vanishes ({\it{i.e.}} $B\rightarrow 0$) we have $P_{||}=P_\perp$ and thus $\Pi_{\mu\nu}\rightarrow 0$ (from (\ref{eq:TabMatricial}), (\ref{PB}) and (\ref{eq:Pzz})--(\ref{eq:Pxx})). As a consequence,  (\ref{Tenab}) reduces to the momentum-energy tensor of a perfect fluid with isotropic pressure:
\begin{equation} T_{\mu\nu}=\rho \, u_\mu u_\nu + P\, h_{\mu\nu}+\Lambda\,g_{\mu\nu},\qquad P=P_{||}=P_\perp .\label{Tab}\end{equation}
However, an isotropic pressure does not imply an isotropic geometry ({\it i.e.} an FLRW geometry), as the energy--momentum tensor (\ref{Tab}) can still be compatible with the inherent anisotropy of the Bianchi I geometry that is present in the different fluid expansion rates in (\ref{eq:def theta}) and the nonzero shear tensor in (\ref{eq:def sigma}) and (\ref{eq: componentes}) (see Appendix \ref{KVariables}). Hence, a zero magnetic field in a Bianchi I model only leads to an FLRW geometry if besides the magnetic field the shear tensor vanishes as well. 

The conditions for an evolution with zero or nonzero shear follow readily from the evolution equations (\ref{EqS2})--(\ref{EqS3}), which if the source is a perfect fluid ($\mathcal{P}_\perp=\mathcal{P}_{||}$ from (\ref{LasGamas})) can be integrated formally as
\begin{equation}
S_1(\tau) = S_1(\tau_i)\exp\left(-3\int{{\mathcal H}\,d\tau}\right),\qquad S_2(\tau) = S_2(\tau_i)\exp\left(-3\int{{\mathcal H}\,d\tau}\right),\label{shear12}\end{equation}
with $S_3=-(S_1+S_2)$. Hence, when $B=0$ and the energy--momentum tensor becomes (\ref{Tab}) we can identify the following two possibilities:
\begin{itemize}
\item Evolution with nonzero shear. If the initial values ($S_{\rm{a}}(\tau_i)$ for $\rm{a}=1,2,3$) of (at least) one of the shear eigenvalues is nonzero the shear tensor is nonzero for all $\tau>\tau_i$. Notice that (for example) $S_2(\tau_i)=0$ implies $S_3(\tau_i)=-S_1(\tau_i)$, which is nonzero in general.
\item Evolution with zero shear: if any two of the initial values $S_{\rm{a}}(\tau_i)$ (for $\rm{a}=1,2,3$) vanishes, then $S_1=S_2=S_3=0$ holds for all $\tau>\tau_i$. The shear tensor vanishes. 
\end{itemize}
We examine below these two limit cases for a zero magnetic field and the possibility of studying the latter in a purely FLRW context.

\subsection{Perfect fluid Bianchi I limit}

If $B=0$ but the shear tensor is nonzero ($S_{\rm{a}}(\tau_i)\ne 0$ for at least one of the $S_{\rm{a}}$)  we have a Bianchi I model whose source is the same particle mixture described in section \ref{sources} but without the magnetic interaction. This source is characterised by the equations of state (\ref{PB}),  (\ref{eq:Pzz})--(\ref{eq:Uep}) with $B=0$, leading for each particle species to    
\begin{eqnarray}P^k = P_{\bot}^{k}= P_{||}^{k} = -\Omega^{k},\qquad  \rho^{k}=-\Omega^{k}+TS+\mu N,\\
P_{\bot}^B=-P_{\|}^B=\rho^B=0,\end{eqnarray}
where now the statistical thermodynamical potential $\Omega^k=\Omega^{k}_{\textrm{\tiny{SQFT}}}(\mu,T)$ in (\ref{Omega Statiticaln}) does not depend on $B$. The remaining equation (\ref{Omega Vn})--(\ref{Omega Statiticaln}) remain valid with $B=0$. The dynamics follows from Einstein's equations, which is the system (\ref{EHt})--(\ref{EqT}) and (\ref{constraint}) with $\mathcal{P}_\perp=\mathcal{P}_{||}=\mathcal{P}$ and without the Maxwell part: {\it i.e.} without (\ref{EqB}) since $\mathcal{B}=0$.

\subsection{FLRW limit model} 
  
If $B=0$ and the shear tensor vanishes ($S_{\rm{a}}(\tau_i)=0$ for at least two of $S_{\rm{a}}$), then $P_\perp=P_{||}=P$ and $S_{\rm{a}}=0$ holds for all $\rm{a}=1,2,3$, hence $a_1=a_2=a_3=a$. It is straightforward to show that  (\ref{eq:Gxx})--(\ref{EQU})  reduce to the well known spatially flat FLRW equations:
\begin{equation}
 \frac{\dot a^2}{a^2}+\frac{2\ddot a}{a}=-\kappa P,\qquad \frac{\dot a^2}{a^2}=\frac{\kappa}{3}\rho,\qquad \dot\rho = -3(\rho+P)\frac{\dot a}{a}, \label{EDinFLRW}\end{equation}
or, in dimensionless form for the particle mixture:
\begin{equation}
 {\mathcal H}_{,\tau} = -\frac{3}{2}\left(\mathcal{E}+\mathcal{P}\right),\qquad \mathcal{T}_{,\tau}=-\frac{3{\mathcal H}\,\left(\mathcal{E}+\mathcal{P}\right)}{\mathcal{E}_{,\mathcal{T}}},\qquad {\mathcal H}^2 =\mathcal{E},\label{EDinFLRW2} \end{equation}
where $\mathcal{P}\equiv P/ \lambda$.

\section{Perturbations, the ``weak field'' approximation and late time evolution}\label{addenda}

Magnetic fields are evidently incompatible with a non-perturbed FLRW Universe, but the expected near homogeneity and high electric conductivity of the early Universe cosmic fluid before recombination \cite{Dario_review,Kandus:2010nw,Widrow:2012:rev,Durrer:2013:rev,Tsagas} suggests introducing the magnetic interaction through suitable scalar and vector perturbations on an FLRW background. Such a perturbative approach of a frozen magnetic field is fully justified if the latter is tangled on scales smaller than the Hubble horizon \cite{Tsagas} and has lead to a comprehensive literature \cite{Tsagas:2007yx,Tsagas,Tsagas:1999ft,Tsagas:1999tu,Clarkson:2002dd,Tsagas2} in which the perturbations are covariant and gauge invariant. 

\subsection{The magnetised Bianchi I model as an ``exact'' perturbation}\label{magperts}

The usage of an exact Bianchi I model to describe a magnetised fluid mixture is not, strictly speaking, a perturbative treatment of the magnetic interaction. However, if we assume the anisotropy of the energy--momentum tensor to be only caused by a magnetic interaction associated with very weak magnetic fields (as required by observational bounds), then the Bianchi I model can roughly approximate a ``near FLRW'' perturbative--like regime that also complies with exact spatial flatness (which to great extent holds in the early Universe cosmic fluid). Considering this regime is well justified, given the fact that deviations from isotropy (in both the matter--energy source and the spacetime geometry) are expected to be negligible in early cosmic times. Evidently, (from  (\ref{rhoTot})--(\ref{PparTot}), (\ref{eq:Pzz})--(\ref{eq:Uep}) and (\ref{eq:func adimen})) a weak magnetic field implies a near perfect fluid: 
\begin{equation} |\mathcal{P}_\perp- \mathcal{P}_{||}| = \frac{|2B^2-8\pi\sum_{k} B\,M^k|}{8\pi\lambda}\ll 1,\end{equation}
which implies in turn (from (\ref{eq:Gxx})--(\ref{eq:Gzz}) and (\ref{eq:def theta})--(\ref{eq: componentes})) approximately equal scale factors close to a unique FLRW scale factor:  $a_1\approx a_2\approx a_3\approx a(t)$, leading 
to  negligible shear $|\Sigma_{\rm{a}}|/H=|S_{\rm{a}}|\ll 1$ and a near isotropic expansion $\mathcal{H}\approx a_{,\tau}/a+O(S_{\rm{a}})$, as well as (from (\ref{EqB})) to an approximate  ``weak field'' scaling $\mathcal{B}_{,\tau}\approx -2\mathcal{B}\mathcal{H}+O(S_{\rm{a}})\,\,\Rightarrow\,\,B\sim a^{-2}$. In fact, it is straightforward to show that under these conditions 
the evolution equations (\ref{EHt})--(\ref{constraint}) become linearised, taking the form of a set of FLRW ``background'' equations (\ref{EDinFLRW2}) plus two extra ``first order'' linear equations for  ${\mathcal B}$ 
and the shear eigenvalues as ``exact'' perturbations (this linearisation of Bianchi I geometry for magnetised sources is discussed in \cite{Tsagas:1999tu}).  

However, while this type of ``exact'' magnetic perturbations can be cast in terms of the variables of the standard gauge invariant or the covariant formalisms, they are extremely restrictive because the homogeneity of the Bianchi model introduces a magnetic field of infinite extent (that permeates the whole 3--dimensional space) and it only allows to examine the (small) deviation from the FLRW background of the time dependent amplitude of the (exactly) perturbed magnetic source, but (lacking spatial dependence) the ``near FLRW'' Bianchi model cannot describe its dependence on spatial scales that would arise naturally in the standard perturbation formalisms. In other words: a magnetised Bianchi I model that becomes arbitrarily close to its FLRW particular case (by considering sufficiently weak magnetic fields) may only describe approximately an exact perturbation of a unique scale of infinite spatial extent, and as such can approximate only very large scale supra--horizon perturbations as, for example, in various string inspired inflationary magneto--genesis scenarios (see details in \cite{Dario_review,Kandus:2010nw,Widrow:2012:rev,Durrer:2013:rev} and also in \cite{weakfield}), and in the Bianchi I model used in \cite{freestream} to study for the interaction of free streaming neutrinos and such large scale weak magnetic field.

\subsection{Magnetic fields in an FLRW context}\label{pure FLRW}

It is more natural to compare the magnetised Bianchi I model that we have examined with previous work in \cite{Dario_review,Dario95,Dario96:1996,Kawasaki12,hortua2013}, where (besides considering a similar theoretical framework based on Satistical Mechanics) the magnetic interaction in the infinite conductivity regime and the cosmic fluid are both studied in a purely FLRW context, in which   the frozen magnetic field is introduced as part of the background (the ``weakly magnetised FLRW'' spacetime in \cite{hortua2013}) as a  relativistic scalar correction to the energy density and the isotropic pressure of a perfect fluid particle mixture considered as source of an FLRW metric (see equations  (3.26)--(3.29) of \cite{Dario_review} and section IX of \cite{hortua2013}). This approach represents a further simplification of the weak field limit that is applicable to an early radiation dominated era  (well before structure formation) in which matter and radiation  perturbations are also neglected, and can be justified if the magnetic field  is  ``not too tangled on scales smaller than the magnetic dissipation scale'' (see discussion in \cite{Matese} and \cite{Dario_review}).  This simplification of the weak field regime is not gauge invariant, but for the cosmic times under consideration it yields the same solutions as the weak field (and thus as gauge invariant perturbations) in the infinite conductivity regime \cite{Tsagas2,weakfield}.  

While the energy density of the perfect fluid mixture described above fully coincides with our total energy density in equation (\ref{rhoTot}), we consider anisotropic pressures in order to be consistent with the full non--perturbative treatment based on Einstein--Maxwell field equations.  Under the approach of \cite{Dario_review,Dario95,Dario96:1996,Kawasaki12} the conservation of the frozen magnetic flux yields:
\begin{equation} \frac{\dot B}{B}+\frac{2\dot a}{a}=0\quad \Rightarrow\quad B(t)=\frac{B(t_i)}{a^2(t)},\label{BFLRW}\end{equation}
which can be compared with the exact  equation (\ref{MEq1}) of the Einstein--Maxwell system obtained in a Bianchi I Universe (or equation (\ref{EqB}) in dimensionless variables):
\begin{equation} \frac{\dot a_{1}}{a_{1}}+\frac{\dot a_{2}}{a_{2}}+\frac{\dot{B}}{B}=0 \quad \Rightarrow\quad B(t) = \frac{B(t_i)}{a_1(t)\,a_2(t)}.\label{MEq2} \end{equation}
While (\ref{BFLRW}) can be regarded as an approximation of (\ref{MEq2}) if anisotropy is  negligible (so that $a_1\approx a_2$), our approach is different from that of \cite{Dario_review,Dario95,Dario96:1996,Kawasaki12}: we regard the Maxwell equation (\ref{MEq2}) as part of the coupled Einstein--Maxwell system in the Bianchi I geometry, whereas the flux conservation (\ref{BFLRW}) in \cite{Dario_review,Dario95,Dario96:1996,Kawasaki12} merely provides a subsidiary condition for Einstein's equations in the FLRW metric and $B$ is taken as a scalar (which is not correct: it is the component of  a frame--dependent magnetic field vector). Another important difference with respect to  \cite{Dario_review,Dario95,Dario96:1996,Kawasaki12} is that these authors simply assume the FLRW radiation temperature for the magnetised mixture, whereas in our Einstein--Maxwell system this temperature needs to be obtained from the evolution equation  
\begin{equation}\label{EQTBI}
\dot{T}=\frac{1}{\rho_{,T}}\left[ -\left(\frac{\dot a_{1}}{a_{1}}+\frac{\dot a_{2}}{a_{2}}\right)\left(P_{\bot}+\rho\right)-\frac{\dot a_{3}}{a_{3}}\left(P_{\|}+\rho\right)+ \left( \frac{\dot{a_{1}}}{a_{1}}+\frac{\dot{a_{2}}}{a_{2}}\right) B \rho_{,B} \right],
\end{equation}
which is coupled with the magnetic field and expressed in terms of dimensionless variables (\ref{LasGamas})--(\ref{eq:func adimen}) becomes equation (\ref{EqT}). However, if we assume an FLRW geometry $a=a_1=a_2=a_3$, a perfect fluid source $P_\perp=P_{||}$ and the radiation equation of state, then (\ref{EQTBI}) becomes the FLRW evolution equation for the radiation temperature: $\dot T=-T\dot a/a$.  Nevertheless, for very weak magnetic fields (\ref{EqT}) should approximate this radiation temperature  evolution law.    

Therefore, the results of \cite{Dario_review,Dario95,Dario96:1996,Kawasaki12} can always be obtained from our Bianchi I based results if we assume the same approximations considered by these authors. While these approximations may be well justified for weak fields in an early time cosmic mixture, our  non--perturbative approach allows us to examine magnetised cosmic fluids also when such assumptions cannot be justified, as we consider the coupled Einstein--Maxwell system that  takes into full account the spacetime anisotropy, its effects on the energy density and pressures and the vectorial (frame--dependent) nature of the magnetic field.

\subsection{Late time evolution}\label{latetimes}

For cosmic times beyond  nucleosynthesis we need to consider the evolution of the tightly coupled fluid mixture in the continuing radiative classical plasma phase before recombination, and then the matter dominated phase after the decoupling of radiation and matter, where a hydrodynamical regime is no longer a valid assumption. For the parameters and initial conditions we have considered (see Sec. \ref{results}) the near FLRW evolution at late times yields baryon and CDM densities  diluting as $\sim a^{-3}$, radiation, neutrino and magnetic field densities as $\sim a^{-4}$, while the cosmological constant dominates the dynamics as the Bianchi I model evolves asymptotically into a $\Lambda$CDM model. 

Whether we account or neglect possible magneto--genesis or some form of magnetic interaction (as discussed in \cite{befrec}, see also \cite{Kandus:2010nw,Widrow:2012:rev,Durrer:2013:rev}) in the lapse between nucleosynthesis and radiation--matter decoupling, the evolution of the components of the source we have examined in the subsequent matter dominated era needs to be examined by kinetic theory, with the contribution of baryons and CDM dominated by their rest mass density and relativistic particle species (photons and neutrinos) undergoing a free streaming evolution. These are precisely the conditions assumed in \cite{freestream} for the a post recombination evolution of a Bianchi I model that fulfils present day observational bounds on the CMB and tends asymptotically to a $\Lambda$CDM model. By imposing the constraints on the magnetic field that yield the same CMB bounds (see Sec. \ref{evolutB}), the Bianchi I model we have examined  effectively exhibits the same late time evolution as the Bianchi I model of \cite{freestream}. In fact, our model also complies with the nucleosynthesis constraints on the abundance of light elements (see Appendix \ref{CBBN}).

\section{Numerical results}\label{results} 
\subsection{Stages of cosmic evolution}\label{EvolMagUni}
We consider for the numerical study of the dynamics of a magnetised Universe the following two periods of the cosmic evolution before the nucleosynthesis:

\begin{itemize}
\item  {\underline {Epoch I: End of the leptonic era:}} $100\, \hbox{MeV} > T > m_e$  ( $200 \gtrsim \mathcal{T} >1$)

The constituents of the cosmic fluid were leptons (neutrinos-antineutrinos and electrons-positrons), baryons (neutrons and protons), photons and cold dark matter. We  will assume besides these particles a homogeneous (time dependent) magnetic field. At $T \simeq 1 \,\hbox{MeV}$ the neutrinos decouple, however for  temperature values such that $T>m_e$, ($m_e\simeq0.5 \, \hbox{MeV}$),  photons, neutrinos and electron-positrons continue with a unique temperature $T_{e^\pm}=T_\gamma=T_\nu \equiv T$.

\item  {\underline{Epoch II: Beginning of the radiation epoch:}} $m_e > T > 0.06 \, \hbox{MeV}$ ($1> \mathcal{T} \gtrsim 0.12$) \cite{Weinberg}-\cite{Rich}

When the temperature drops below $m_e$, the electron-positron pairs are transformed into photons but not into decoupled neutrinos. After these
annihilations the number of photons is therefore greater than the number of neutrinos.
Since thermal equilibrium is maintained until very few electrons remain, the entropy of the photon-electron-positron system for $T> m_e$ is nearly the entropy of the photon system for $T< m_e$. Thus, after the electron-positron annihilation the photon and neutrino temperatures are related by:
\begin{equation} \label{Tnu}
T_\nu=\left(4/11\right)^{1/3} T_\gamma \,.
\end{equation}
\end{itemize}
Since $\tau \simeq 0.029\, \hbox{s}^{-1} \times (t-t_i)$, the cosmic times for these epochs range from an initial time at the outset of the leptonic era $t_i \simeq 1\,\hbox{s} \Rightarrow \tau_i =0$ at temperature $T=100\, \hbox{MeV}$, towards the final stage at the beginning of nucleosynthesis at $t\simeq 200\,\hbox{s} \Rightarrow \tau \simeq 6$. 

Having defined the cosmic eras we are interested in studying, we undertake in this section the numerical integration of the system (\ref{EHt})--(\ref{EqT}) and (\ref{constraint}) under initial conditions specified at an initial time $\tau=\tau_i$ and following the assumptions summarised below:
\begin{itemize}
\item  An initial temperature of $\mathcal{T}(\tau_i)\simeq200$  and ending our analysis when temperature drops below  $\mathcal{T}(\tau_i)\simeq0.12$.
\item Initial magnetic field is
$0\leqslant\mathcal{B}(\tau_i)\lesssim\mathcal{T}^2(\tau_i)$ (in Gauss 
$0 \leqslant B(t_i)\sim 5\times 10^{17 }\hbox{G}$). 
\item Initial shear is zero: $S_1(\tau_i) = S_2(\tau_i) =  S_3(\tau_i) = 0$. From (\ref{EqS2})--(\ref{EqS3}) and (\ref{shear12}) 
this choice implies that all the anisotropy introduced by a Bianchi I geometry can be ascribed exclusively to the 
magnetic interaction through $\mathcal{P}_\bot\ne \mathcal{P}_\|$. In other words, we have a magnetised Bianchi 
Universe that approaches an FLRW as the magnetic 
field becomes weak or negligible, while the strict limit $B=0$ ($\mathcal{B}=0$) yields a pure FLRW evolution 
(not a perfect fluid Bianchi I model). While the shear tensor in early cosmic times should be absolutely negligible but not strictly 
(mathematically) zero, the assumption $S_{\rm{a}}(\tau_i)=0$ for $\rm{a}=1,2,3$ is a reasonable approximation. 
\item The magnetic field modifies cosmic dynamics by both, the pure magnetic field contribution (Maxwell term) and its inclusion in the statistical treatment of the electron-positron gas. While the Maxwell term is present during the whole evolution, the electron-positron gas is considered as a magnetised gas
only during the leptonic era.
\end{itemize}

\subsection{Kinematics}

\begin{figure*}[t]
      \begin{minipage}[t]{0.6\linewidth}
      \includegraphics[width=1.0\textwidth]{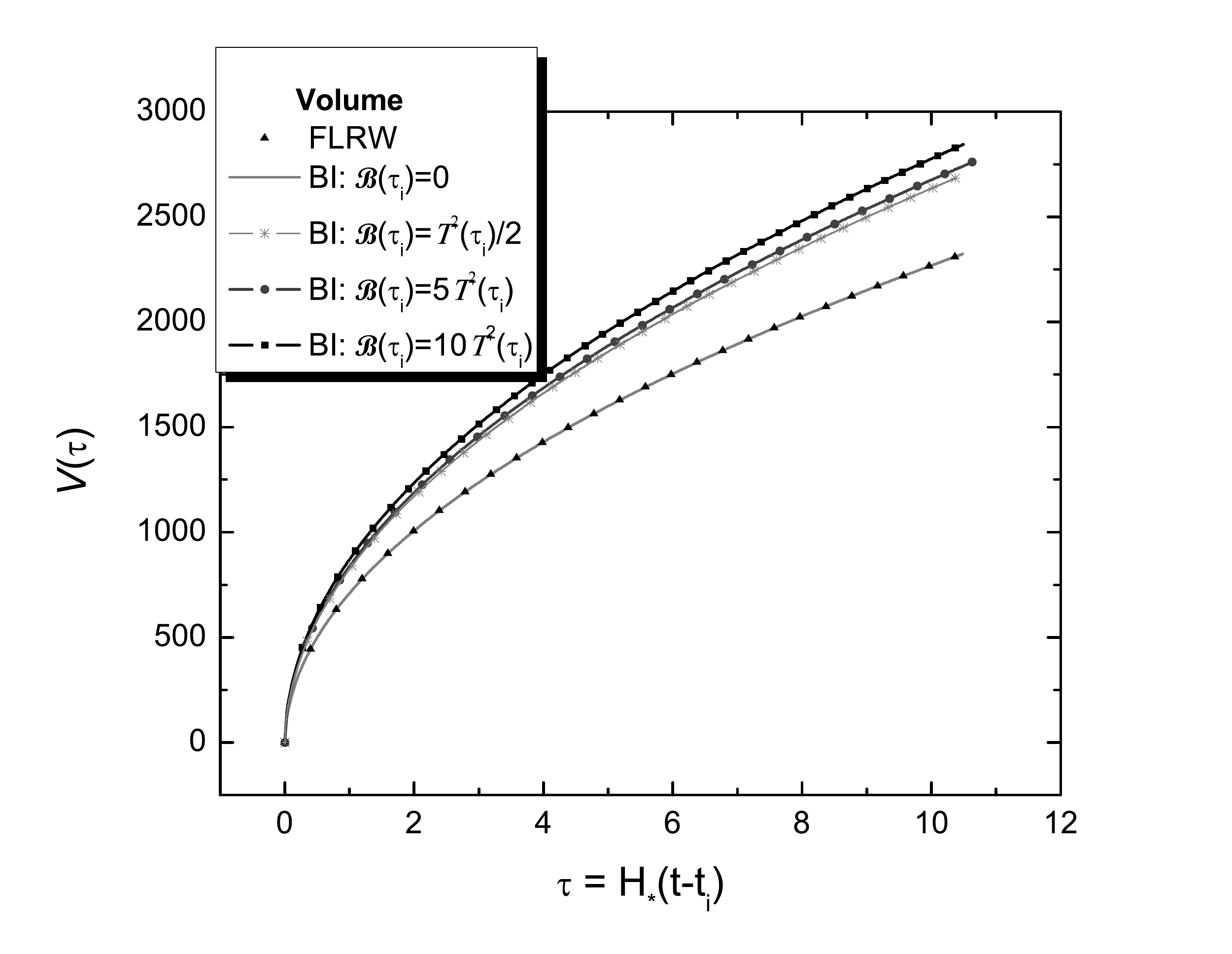}
      \end{minipage}
      \hspace{0.5cm}
      \begin{minipage}[t]{0.65\linewidth}
      \includegraphics[width=1.0\textwidth]{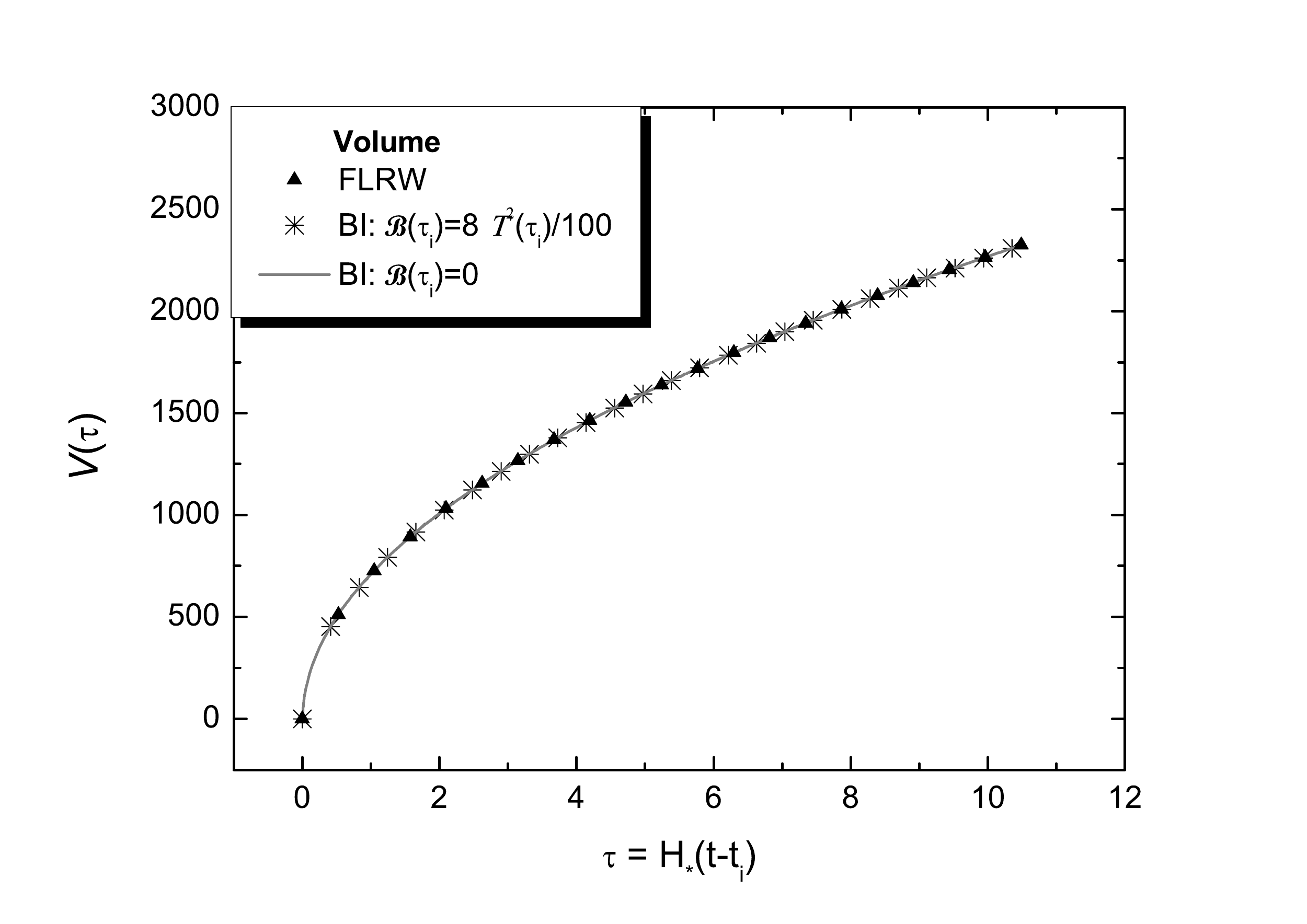}
      \end{minipage}
       \caption{ Proper volume as function of time for increasing values of the initial magnetic field. The panel in the left hand side shows that larger values of $ \mathcal{B}(\tau_i)$ produce faster expansion rates, while the curves of the panel in the right hand side show that the expansion rate of $V(\tau)$ is practically indistinguishable from an FLRW expansion if we consider  values of $ \mathcal{B}(\tau_i)$ that fit observational constraints \cite{Dario96:1996}.}\label{Comparison_BIvsRW}
\end{figure*}

\begin{figure}
\begin{center}
\includegraphics[scale=0.35]{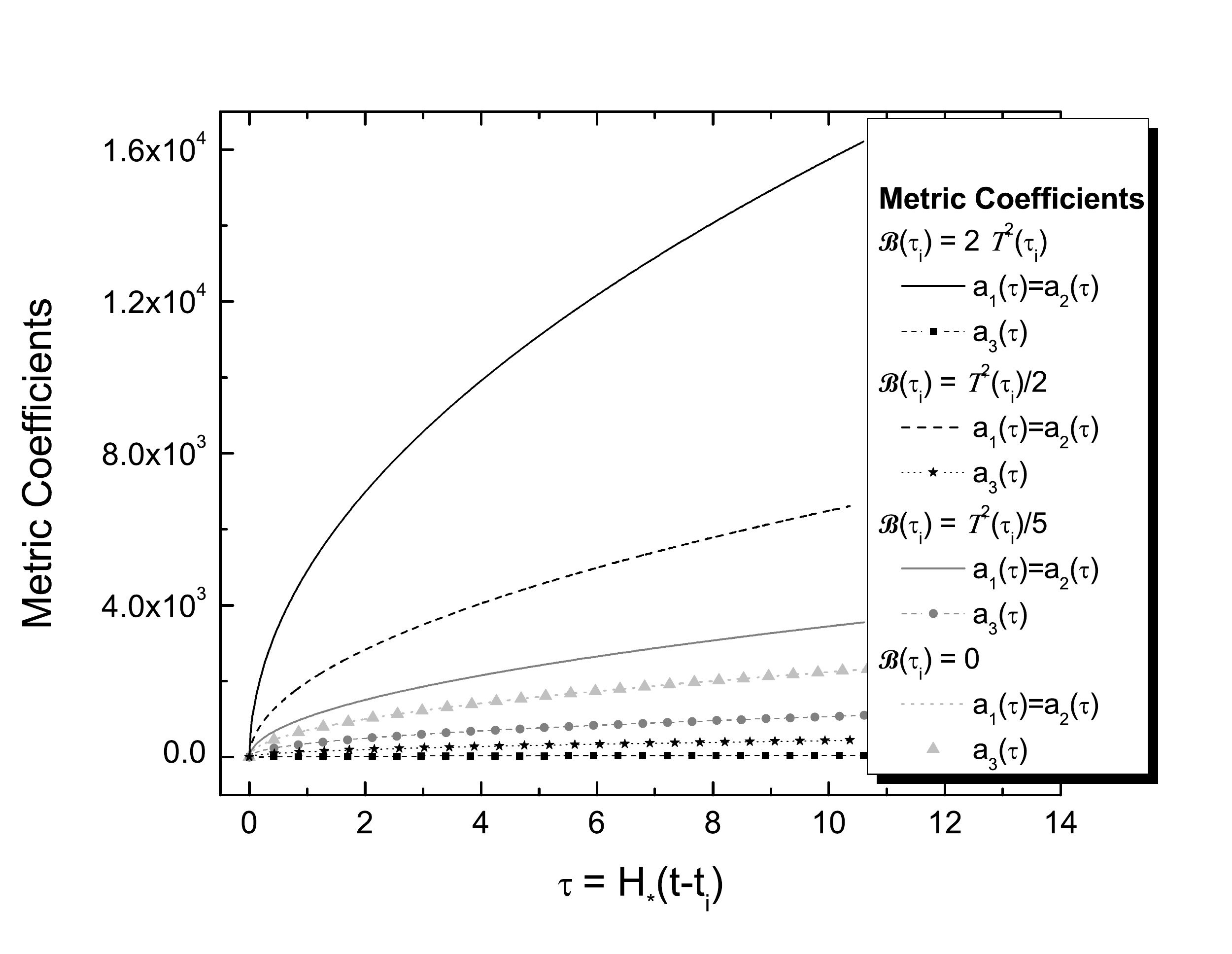}
\caption{Metric coefficients $a_1(\tau),\,a_2(\tau),\,a_3(\tau)$ for varied values of $ \mathcal{B}(\tau_i)$. Larger values of $ \mathcal{B}(\tau_i)$ lead to very different curves (large anisotropy) for the three scale factors, while smaller values  lead to almost the same curves for them (almost isotropy).}
\label{QSb2T2}
\end{center}
\end{figure}

We examine in figure \ref{Comparison_BIvsRW} the local proper volume for different initial values $\mathcal{B}(\tau_i)$ of the magnetic field. The figure reveals a faster rate of expansion for larger $\mathcal{B}(\tau_i)$. However, if we consider values of $\mathcal{B}(\tau_i)$ compatible with observational bounds \cite{Dario96:1996}, then the volume expansion is practically coincident with the expansion rate of the FLRW model obtained when we set $\mathcal{B}(\tau_i)=0$ (see the grey line and the black triangles in the enclosed graph). In other words, the anisotropy effects on the volume expansion are completely negligible when the magnetic field complies with observational bounds. 
Since we are assuming zero initial shear, the magnitude of the initial magnetic field intensity $\mathcal{B}(\tau_i)$ should determine the difference of the expansion rates for the three metric coefficients $a_1,\,a_2,\,a_3$. This  can be appreciated  in figure \ref{QSb2T2} which displays these functions for various values of $\mathcal{B}(\tau_i)$, showing  distinct curves for these metric functions ({\it i.e.} larger anisotropy) for larger values of $\mathcal{B}(\tau_i)$  and  an almost isotropic FLRW expansion (almost the same evolution for the three scale factors) for a weak field.  
\subsection{Thermodynamics}
 The anisotropy inherent in the magnetic field can also be appreciated through the anisotropy of the pressures. The classical Maxwellian contribution $\mathcal{B}^2$ produces  a decelerated fluid expansion in the direction of the magnetic field ($z$), as the latter produces a negative pressure or tension in its direction, and an accelerated expansion in the directions perpendicular to it due to a positive pressure.  However, we obtain the opposite effect from the statistical contribution of $\mathcal{B}$ in the equation of state for the  magnetised electron--positron gas mixture: the pressure in the direction parallel  to the field increases and that in the perpendicular direction decreases  \cite{Chaichian:1999gd,Martinez:2003dz,Gaspar:2013eca,Can1}.  Although the dominant effect  in  the dynamics comes from the contribution $ \mathcal{B}^2$, we can observe that including the 
magnetised electron--positron gas leads to a slight acceleration of the cosmic rate of expansion. In fact, the contribution of the 
electron--positron gas is of the order of $O(\alpha) \times \mathcal{B}^2$ (where $\alpha$ is the constant of fine structure)\cite{Vachaspati}, 
therefore this contribution will always  be sub-dominant in comparison with that of the ``pure'' magnetic field $\sim  \mathcal{B}^2$.
 
\begin{figure}
\begin{center}
\includegraphics[scale=0.35]{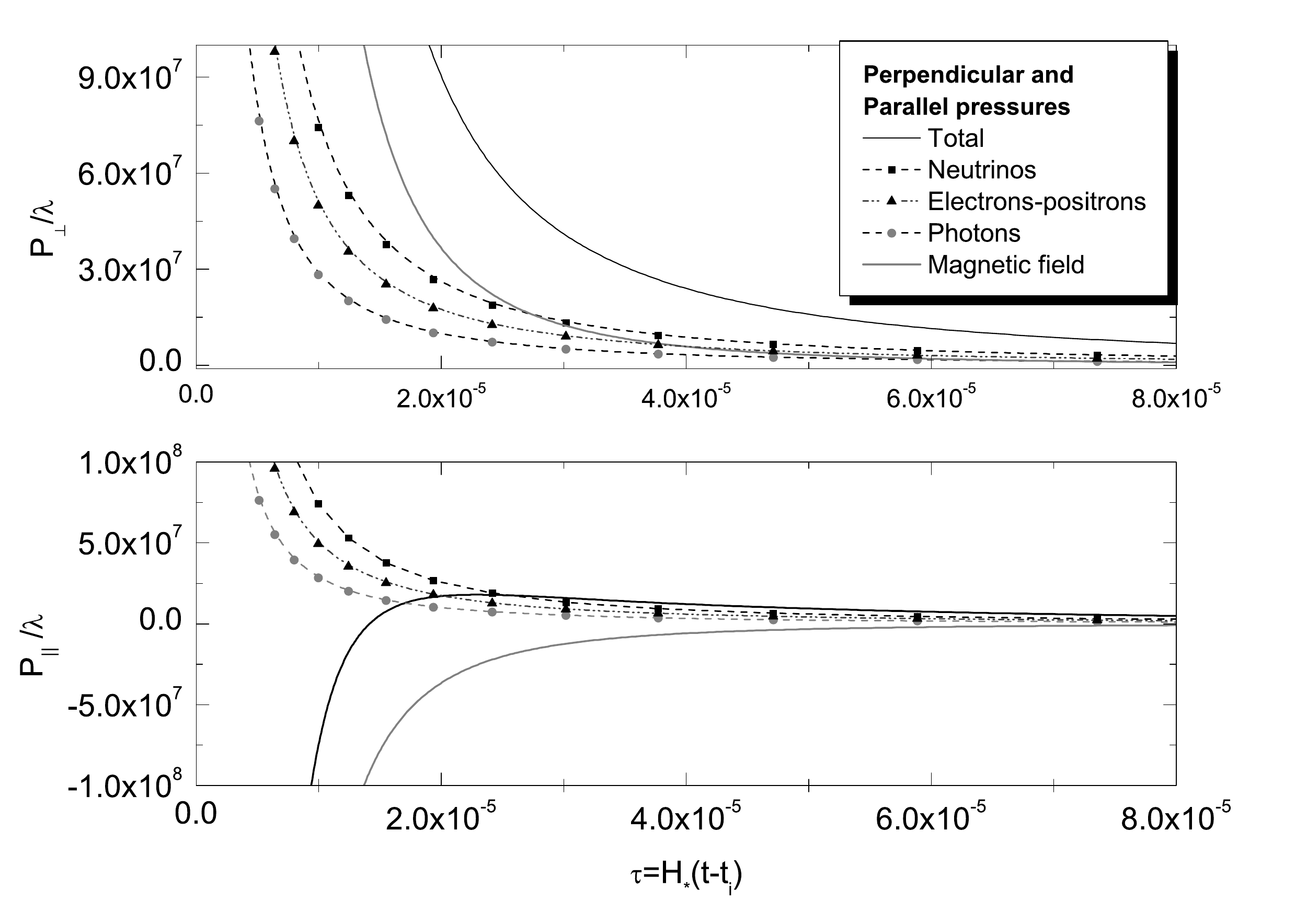}
\caption{Perpendicular and parallel pressures for each mixture component. We assumed $\mathcal{B}(\tau_i)=2 \times \mathcal{T}^2(\tau_i)$. See further explanation in the text.}
\label{Ps}
\end{center}
\end{figure}

\begin{figure}
\begin{center}
\includegraphics[scale=0.35]{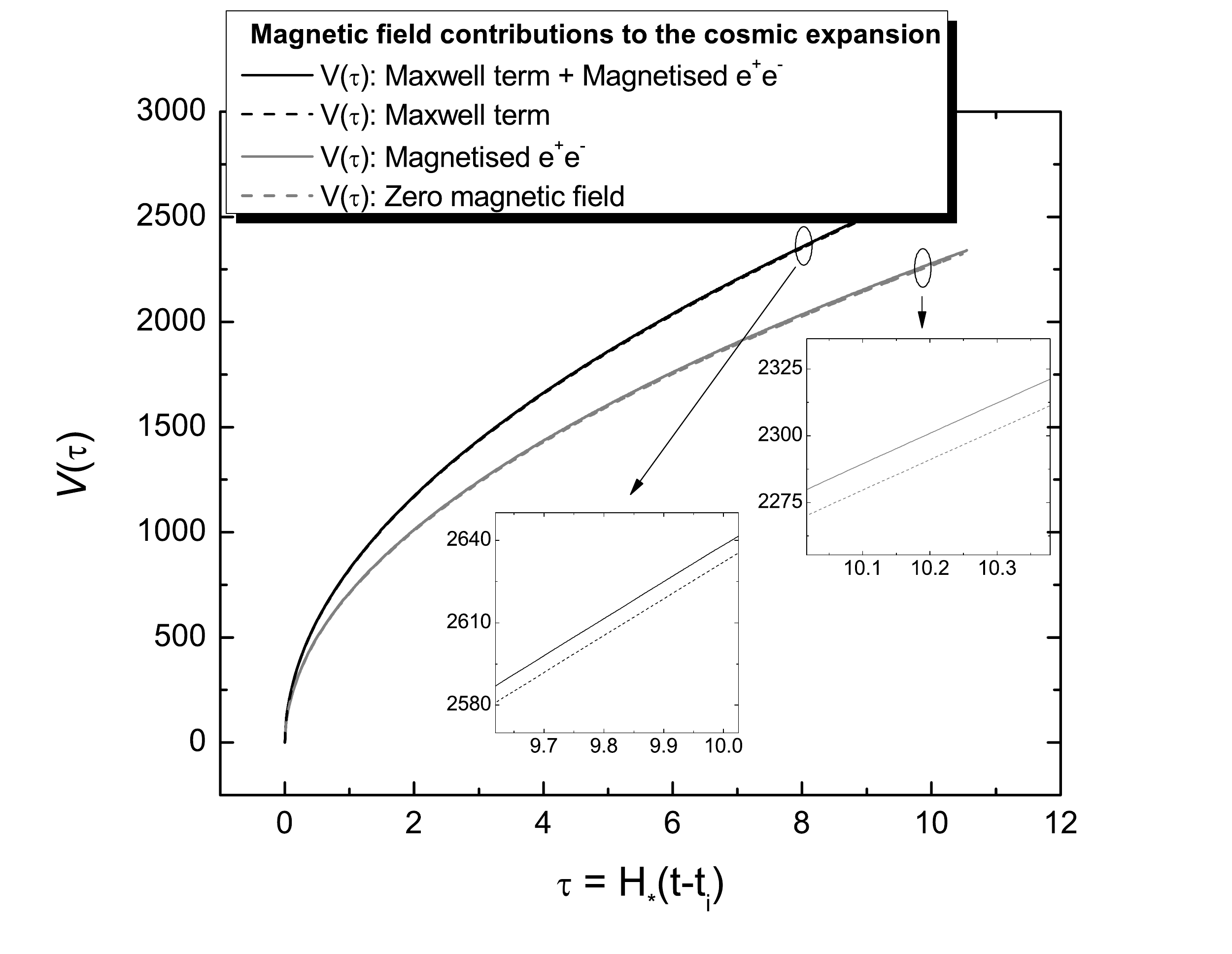}
\caption{Contribution of each magnetic field effect in the volume expansion. We assumed $ \mathcal{B}(\tau_i)=2 \times \mathcal{T}^2(\tau_i)$.}
\label{V_BI}
\end{center}
\end{figure}

The dominance of the Maxwellian contribution $\mathcal{B}^2$  in the fluid expansion is consistent with the curves displayed in figure \ref{QSb2T2}: the scale factor $a_3(\tau)$ (direction parallel to $\mathcal{B}$) expands at a slower rate than $a_2(\tau)$ and $a_1(\tau)$ (directions perpendicular to $\mathcal{B}$). This dominance is also displayed in figures \ref{Ps} and \ref{V_BI}, where we considered magnetic field values much larger than observational bounds to highlight this effect. Figure \ref{Ps} shows that the pressure $\mathcal{P}_\bot$ is much smaller than  $\mathcal{P}_\|$, while figure \ref{V_BI} depicts the growth of the volume expansion for different interactions of the magnetic field:  the electron--positron as the magnetised/(\emph{non--magnetised}) gas (solid /(\emph{dashed}) curves) and the grey curves (solid or dashed) represent the volume expansion  obtained after eliminating the Maxwell term in the pressures and energies.

\begin{figure}
\begin{center}
\includegraphics[scale=0.35]{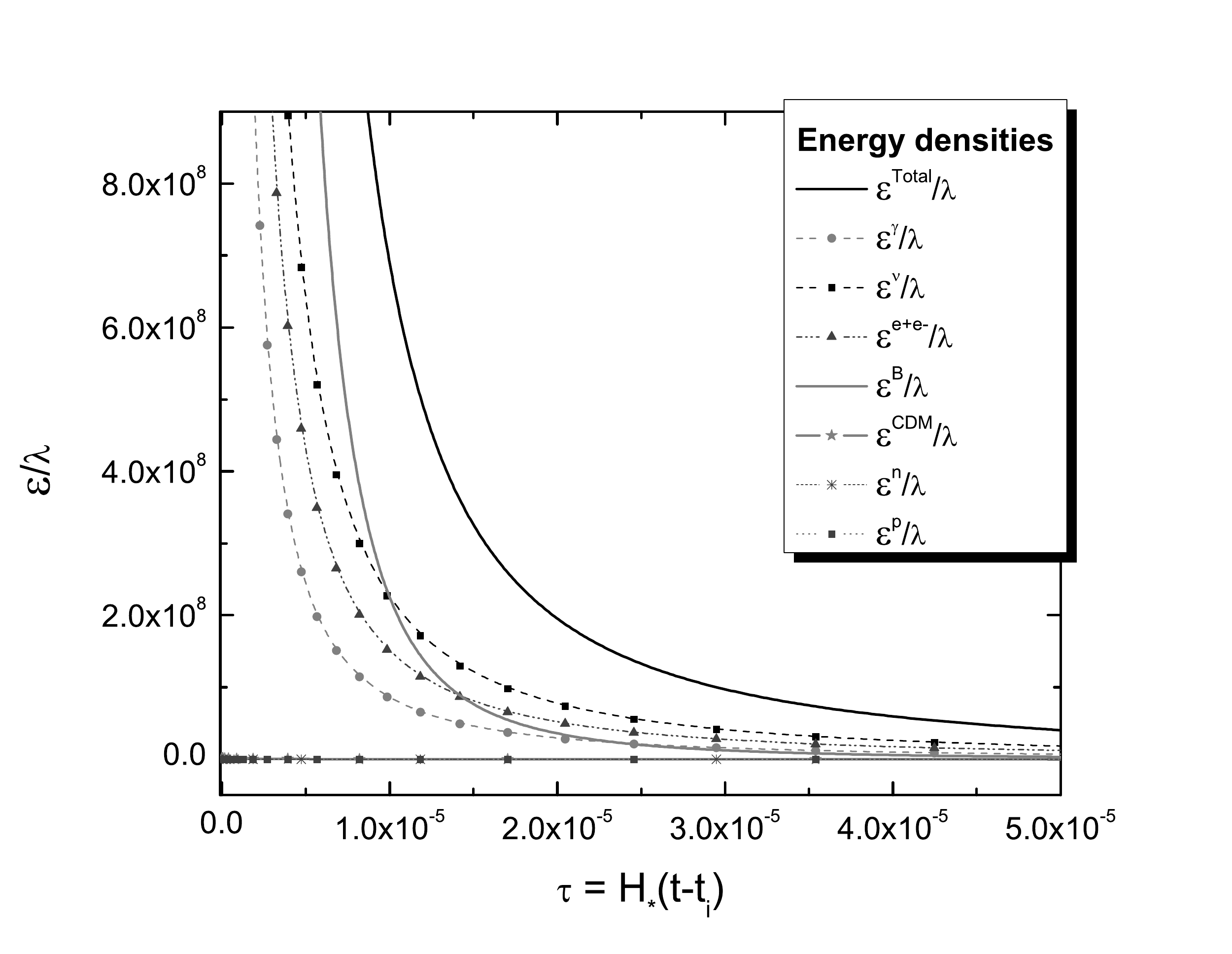}
\caption{Energy densities (total and for each mixture component). We assumed $ \mathcal{B}(\tau_i)=2 \times  \mathcal{T}^2(\tau_i)$.}
\label{UI}
\end{center}
\end{figure}

So far, we have not mentioned the contribution of the magnetic field  to the energy density and to the fluid dynamics. This contribution depends on its initial value $\mathcal{B}(\tau_i)$. It is negligible when  $\mathcal{B}(\tau_i)\ll \mathcal{T}^2(\tau_i)$  but becomes  significant when  $\mathcal{B}(\tau_i) \sim \mathcal{T}^2(\tau_i)$. In the former case   $\mathcal{B}$ and $\mathcal{T}$ exhibit almost identical behaviour to that of the well known FLRW solutions (along the lines of  \cite{Dario_review}) that follow by  assuming $\mathcal{T} \sim 1/a$ and $\mathcal{B}\sim 1/a^2$, where $a$ is the FLRW scale factor for a radiation dominated fluid.  However, if $\mathcal{B}(\tau_i) \sim \mathcal{T}^2(\tau_i)$ there are significant differences with the FLRW evolution. In order to explore these differences we depict in Figure \ref{UI} the energy density for the different components of the fluid mixture during the leptonic era, assuming an (unrealistic) initial magnetic field $\mathcal{B}(\tau_i)=2 \times \mathcal{T}^2(\tau_i)$ that is much larger than values allowed by observational bounds. 
As follows from these graphs, the curve that corresponds to the pure magnetic Maxwell term contribution ($\propto \mathcal{B}^2$) decays very fast, but is overtaken by the curves for the  neutrinos, the electrons--positrons and the photons. This behaviour is very different from the expected radiation--like FLRW behaviour $\mathcal{T} \sim 1/a$ and $\mathcal{B}\sim 1/a^2$ of much weaker magnetic fields (subjected  to observational constraints). Hence, if $\mathcal{B}(\tau_i) \sim \mathcal{T}^2(\tau_i)$ the Bianchi I and FLRW dynamics lead to very different decay rates  for both the temperature and the magnetic field (see Sec. \ref{LimitRW}, in particular equations (\ref{MEq2})-(\ref{EQTBI})).

\subsection{The magnetic field}\label{evolutB}

\begin{figure*}[t]
      \begin{minipage}[t]{0.61\linewidth}
      \includegraphics[width=1.0\textwidth]{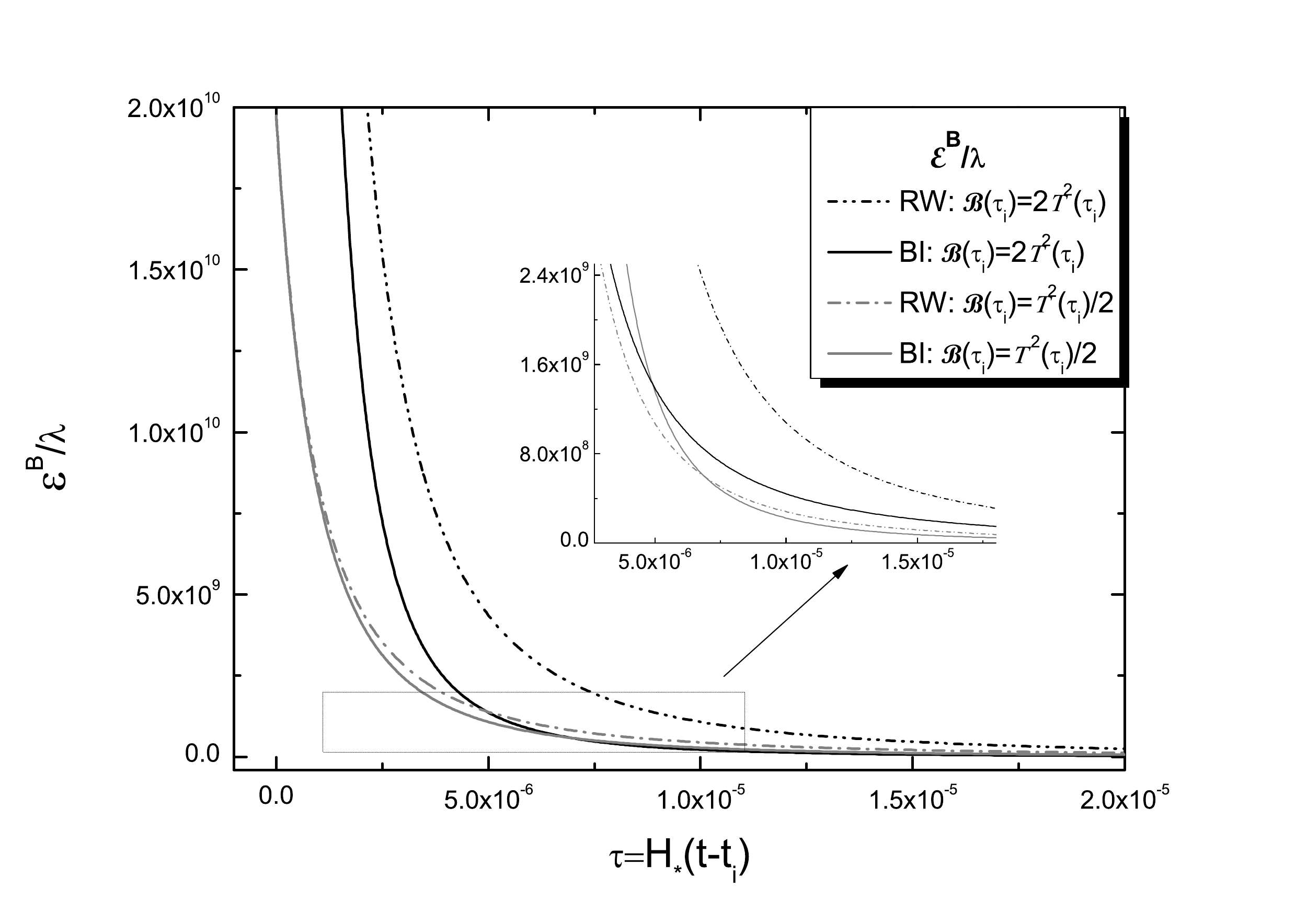}
      \end{minipage}
      \hspace{0.5cm}
      \begin{minipage}[t]{0.60\linewidth}
      \includegraphics[width=0.82\textwidth]{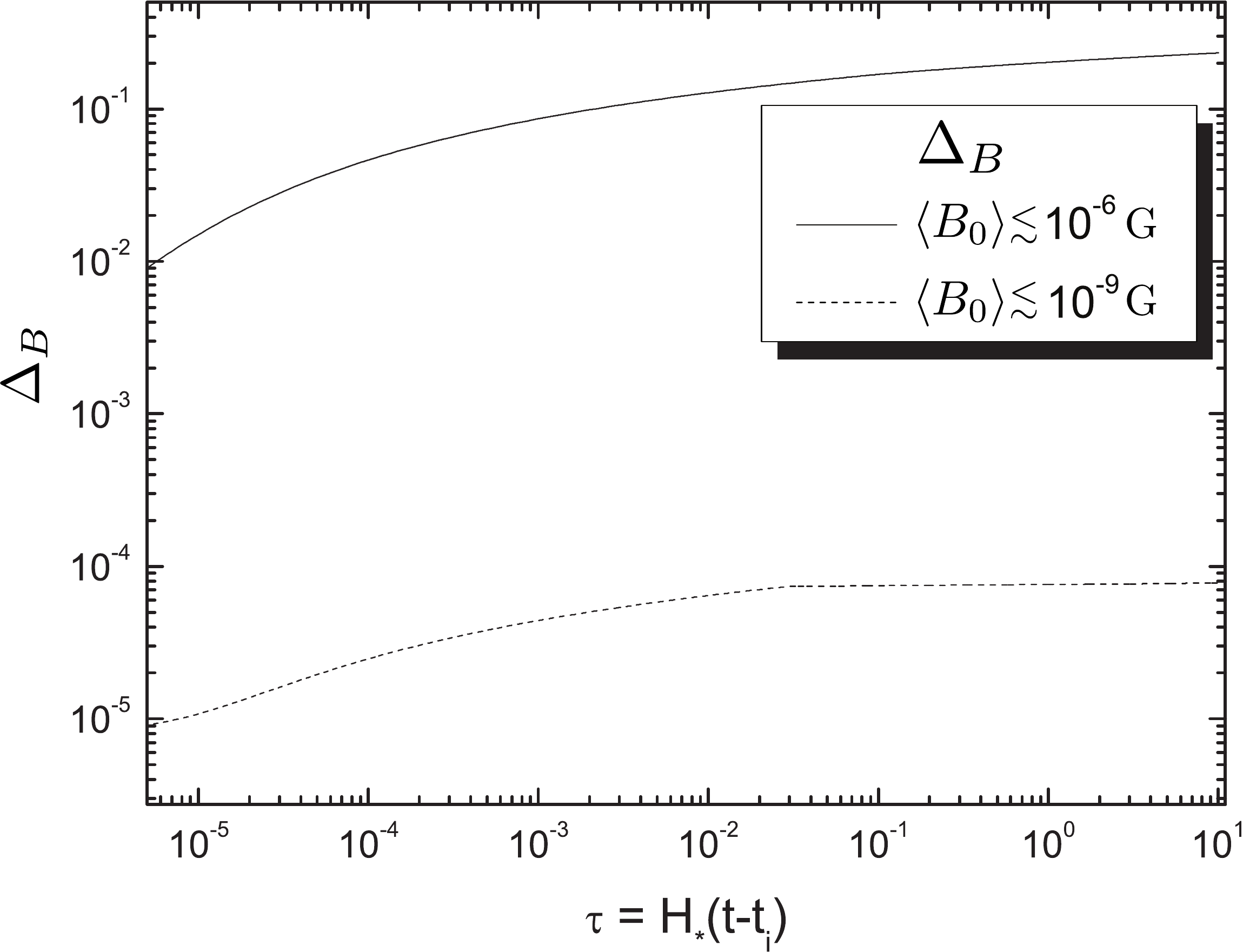}
      \end{minipage}
       \caption{ Pure magnetic energy in the magnetised Bianchi I model and in a FLRW context (assuming $\mathcal{B}\propto a^{-2}$) for various values of $ \mathcal{B}(\tau_i)$ (left hand side panel). 
Relative error: $\Delta_\mathcal{B}(\tau)=[\mathcal{B}(\tau)-\mathcal{B}(\tau_i)/a^2 (\tau)]/\mathcal{B}(\tau)$ is plotted in the right hand side 
panel for magnetic fields complying $\left< B_0 \right>\lesssim10^{-6} \hbox{G}$  (solid line) and $\left< B_0 \right>\lesssim10^{-9} \hbox{G}$  (dashed line).
}\label{UB}
\end{figure*}

The different behaviour of the pure magnetic energy density ($ \propto \mathcal{B}^2$) is illustrated in figure \ref{UB}. The solid curves represent the solutions obtained from Bianchi I dynamics, whereas the dashed curves correspond to a magnetic field in an FLRW context ($B\propto 1/a^2$). In both cases we plotted in the left hand side panel curves for two values of the initial magnetic field: $\mathcal{B}(\tau_i)=2 \mathcal{T}^2(\tau_i)$ (black line) and $ \mathcal{B}(\tau_i)=\mathcal{T}^{2}(\tau_i)/2$ (grey line). 
In the right hand side panel we plotted the relative error $\Delta_{\mathcal{B}}(\tau)$ between $\mathcal{B}(\tau)$ obtained from Bianchi I 
 dynamics and its FLRW equivalent $\mathcal{B}(\tau_i)/a^2(\tau)$ for weak magnetic fields complying with
$\left< B_0 \right>\lesssim10^{-6} \hbox{G}$ (solid line) and $\left< B_0 \right>\lesssim10^{-9} \hbox{G}$  (dashed line).
 As shown by this graph, $\Delta_{\mathcal{B}}(\tau)$ increases as the fluid evolves. 
 This is an interesting result: while the evolution of kinematic and state variables for initial values ${\mathcal{B}}(\tau_i)$ of weak fields 
is practically indistinguishable in Bianchi I and FLRW dynamics, the evolution of the magnetic field itself can reveal non--negligible differences. The explanation for this effect is straightforward: while $a^2\approx a_1a_2$ holds for weak fields, each Bianchi I scale factor is related to the FLRW scale factor by a small (but time dependent) correction: $a_1(\tau) \simeq a(\tau)+\epsilon_1(\tau)$ and $a_2(\tau) \simeq a(\tau)+\epsilon_2(\tau)$, hence we have:
\begin{equation}\left[\mathcal{B}\right]_{\textrm{\tiny{BI}}} \approx \left[\mathcal{B}\right]_{\textrm{\tiny{FLRW}}}-\frac{(\epsilon_1(\tau)+\epsilon_2(\tau))\mathcal{B}_i}{a^3(\tau)},\label{FLRWBI}\end{equation}
where $\left[\mathcal{B}\right]_{\textrm{\tiny{FLRW}}}$ and $\left[\mathcal{B}\right]_{\textrm{\tiny{BI}}}$ are the FLRW and Bianchi I scaling laws given by (\ref{BFLRW}) and (\ref{MEq2}). Since we are assuming that the shear tensor vanishes at $\tau=\tau_i$, then $\epsilon_1(\tau_i)=\epsilon_2(\tau_i)=0$, and thus the Bianchi I and FLRW forms initially coincide. However, as the expansion proceeds the ``error'' introduced by $\epsilon_1(\tau)+\epsilon_2(\tau)$ in (\ref{FLRWBI}) (which enters in $\Delta_{\mathcal{B}}(\tau)$) will be negligibly small only for $\tau\approx\tau_i$ but necessarily grows to small (but not necessarly negligible values) as the fluid expands. 
Although the relative error remains negligible for the calculations using the more restrictive bound ($\left< B_0 \right>\lesssim10^{-9} \hbox{G}$), $\Delta_{\mathcal{B}}$ is of the order of $10^{-4}$, it and can reach values of the order of $10^{-1}$,
if we consider $\left< B_0 \right>\lesssim10^{-6} \hbox{G}$.
On the other hand, since even with these corrections the field itself is very weak, the evolution of the kinematic and state variables is practically insensitive to them. 
%

\begin{figure}
\begin{center}
\includegraphics[scale=0.35]{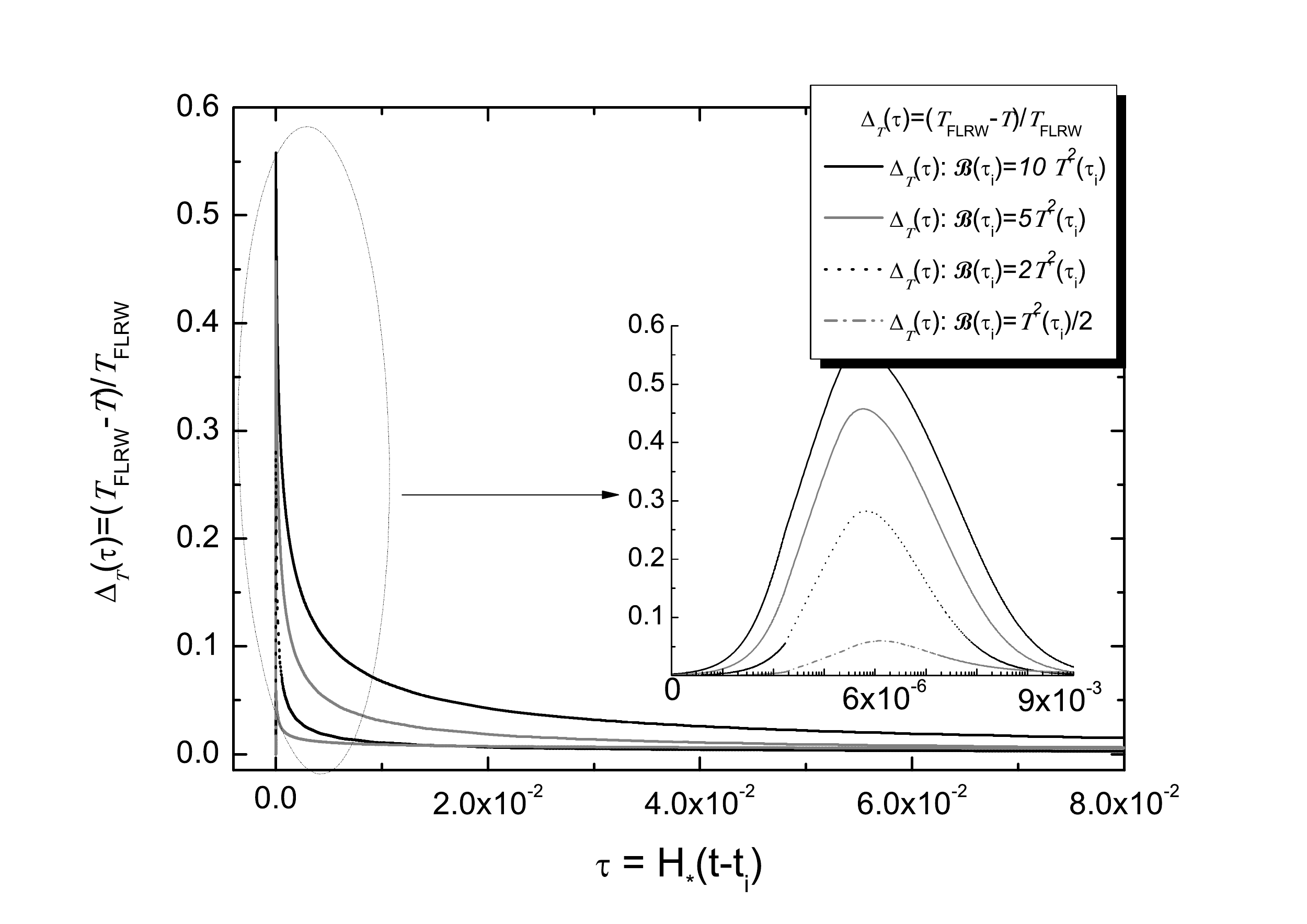}
\caption{The figure displays the relative ratio $\Delta$ defined in (\ref{Delta}) between the temperature for a magnetised Bianchi I fluid mixture and the same fluid with zero magnetic field and an FLRW metric. }
\label{TvsB}
\end{center}
\end{figure}

Finally, we remark that larger initial values of the magnetic field affect the dynamics of the other thermodynamical functions. Since the latter depend on the temperature, this can be illustrated by its different evolution with and without magnetic field. Hence, we depict in figure \ref{TvsB} the evolution of the quantity
\begin{equation}\Delta_{\mathcal{T}}=\frac{\mathcal{T}_{\textrm{\tiny{FLRW}}} -\mathcal{T}}{\mathcal{T}_{\textrm{\tiny{FLRW}}}}\quad\left(=\frac{T_{\textrm{\tiny{FLRW}}} -T}{T_{\textrm{\tiny{FLRW}}}}\right),
\label{Delta}\end{equation} 
which  provides an estimation of the relative difference between the temperature in the magnetised Bianchi I mixture  and the temperature in the non--magnetised FLRW model ($\mathcal{T}_{\textrm{\tiny{FLRW}}}$) that results by setting $\mathcal{B}=0$. 
Since the initial temperature value is fixed, all curves depicted in the figure start at $\Delta_{\mathcal{T}}=0$. As $\tau$ advances the evolution of the temperature of the magnetised mixture differs slightly (depending on the value of $\mathcal{B}(\tau_i)$) from the FLRW evolution without magnetic field, with $\Delta_{\mathcal{T}}$  reaching a maximum value close to $\tau_i$ (when the magnetic field is stronger) and dropping  as the expansion proceeds and the magnetic field decays. However, for weak fields values compatible with observational bounds $\Delta_{\mathcal{T}} \sim O(10^{-5})$,  therefore $\mathcal{T}$ is practically indistinguishable from $\mathcal{T}_{\textrm{\tiny{FLRW}}}$.

\section{Conclusions}\label{conclusion}
We have examined the evolution of a magnetised cosmic fluid mixture between the leptonic and  nucleosynthesis eras by means of a non-perturbative approach to the Einstein--Maxwell field equations. For this study we considered a Bianchi I model, as it provides one of the simplest geometries that is compatible  with the inherent anisotropy of the magnetic field.  We considered as field sources a mixture of baryons (neutrons and protons), leptons (neutrinos, antineutrinos, electrons and positrons), cold dark matter WIMP's  and photons, together with an already existing time dependent magnetic field directed along the $z$-axis (we also assumed an nonzero cosmological constant). For the dynamical study of this cosmic fluid we transformed  the Einstein--Maxwell equations into a first order system amenable for numerical integration.

The general features of the behaviour of the physical quantities during the two epochs we are concerned can be summarised as follows. Temperature, magnetic field and thermodynamical functions depending on them, like pressures and energy densities of each species, start decaying at a fast rate and later at slower rates as the expansion proceeds.  In epoch I at the end of  the leptonic era the neutrinos provide the dominant contribution to the energy density and pressures, followed by the contribution of the electron--positron gas and photons. On the other hand, after the electron--positron annihilation the radiation dominates cosmic dynamics.  These results  agree with well known previous results \cite{Weinberg,Rich}.  

We  have found that cosmic dynamics in the epochs described above is modified by both, the Maxwell term ($\sim B^2$) and the presence of the magnetic field in the equations of state of a magnetised electron--positron gas.
For weak magnetic fields compatible with observational constraints these modifications are negligible (though this is not true for the evolution of the field itself). For magnetic  
fields complying with $eB(t_i)\sim T^2(t_i)$ the main contribution to the dynamics comes from the Maxwell term, leading to an anisotropic expansion that can be appreciated by the different time growth of the metric coefficients and anisotropic pressures. The anisotropy caused by the Maxwell term produces an accelerated expansion in the direction perpendicular to the field and a decelerated one in the field direction. On the other hand, 
the dependence on $B$ in the electron--positron equation of state
acts as a
small correction $O(\alpha)\times \mathcal{B}^2$ to the energy density where $\alpha$  is the fine structure constant (see further detail in \cite{Vachaspati}).

We found that the effect of weak magnetic fields compatible with observational bounds on the evolution of the main thermodynamical variables are negligible.  For such magnetic fields the time decay of these variables is practically undistinguishable from their evolution under a pure FLRW  dynamics. Since observational bounds yield a ``near FLRW'' Bianchi I model that is exactly spatially flat, it resembles some sort of ``exact'' perturbation whose evolution seems to be comparable with that of linear perturbations of magnetised sources on a spatially flat FLRW background \cite{Tsagas:1999tu}. However, because of its homogeneity, the magnetic field in the ``near FLRW'' Bianchi I model permeates the whole time slices and thus cannot describe the scale dependence of standard cosmological perturbations, and thus cannot be used (even approximately) to examine early Universe magneto--genesis that involve seed fields undergoing a scale dependent amplification process (as for example the inverse cascade mechanism or cosmic plasma instabilities) to account for observed magnetic fields of astrophysical interest (see 
\cite{Kandus:2010nw,Widrow:2012:rev,Durrer:2013:rev,invcasc,weibel,befrec}). The only possibility (which was considered in \cite{freestream}) is to regard the ``near FLRW'' magnetised Bianchi I model as providing an approximated description of magnetic fields coherent in very large supra--horizon scales, which may be compatible with some inflationary magneto--genesis scenarios (see \cite{Kandus:2010nw,Widrow:2012:rev,Durrer:2013:rev,non-conformal}).  

Given the limitations of the Bianchi I model mentioned above, we have provided a comparison of our model with the approach of  \cite{Dario_review,Dario95,Dario96:1996,Kawasaki12}, who considered a similar Stetistical Mechanics framework and examined magnetic fields during nucleosynthesis
under  a purely FLRW framework based on the assumption that the spacetime geometry is not affected by magnetic fields in these cosmic times (hence they rule out the anisotropy in the momentum-energy tensor that includes the magnetic interaction). While the results of \cite{Dario_review,Dario95,Dario96:1996,Kawasaki12} can always be obtained from our approach under these same assumptions (which may be well justified for weak fields in a realistic cosmic fluid), our non--perturbative methodology allows us to examine  magnetised cosmic fluids also under conditions in which such assumptions may not be justified.  

A natural continuation of this study is revisiting the magnetic field constraints as well as the study of abundance of light elements, all under the framework of Bianchi I dynamics applied specifically to those magneto--genesis scenarios compatible with supra--horizon scale magnetic fields. This will be attempted in future work.

%


\begin{acknowledgements}
The work of A.P.M and I.D.G has been supported by \emph{Ministerio
de Ciencia, Tecnolog\'{\i}a y Medio Ambiente} under the grant CB0407  and by the ICTP Office of External Activities through NET-35.  A.P.M. also acknowledges the hospitality and support given by the International Center for Relativistic Astrophysics Network where part of this paper was developed.
G.P. aknowledges support from an UNAM-DGAPA-PASPA grant and the hospitality from \emph{Instituto de Cibern\'etica, Matem\'atica y F\'isica, La Habana, Cuba} and from \emph{Depto. de F\'isica Te\'orica y del Cosmos, Facultad de Ciencias, Universidad de Granada, Spain} where this work was conceived and developed.
R.A.S. and I.D.G. acknowledge support from the research grant DGAPA PAPIIT IA101414. 
G.P., A.P.M. and I.D.G. have also received support from UNAM-DGAPA-PAPIIT under grants number
IN117111 and IN117914.
%
\end{acknowledgements}

\appendix
\section{Local kinematic variables}\label{KVariables}

The kinematics of local fluid elements can be described through covariant objects defined by the 4--velocity field $u^\alpha$.
For a Kasner metric in the comoving frame endowed with a
normal geodesic 4-velocity, the only non-vanishing
kinematic parameters are the expansion scalar, $\theta$,
and the shear tensor $\sigma_{\alpha\beta}$:
\begin{equation} \theta=u^{\alpha}\,_{;\alpha}\,,\quad \sigma_{\alpha\beta}=u_{(\alpha;\beta)}-\frac{\theta}{3}h_{\alpha\beta}\,,\end{equation}
where $h_{\alpha\beta}=u_{\alpha}u_{\beta}+g_{\alpha\beta}$ is
the projection tensor and rounded brackets denote symmetrization.
For the Kasner metric these parameters take the form:
\begin{equation}
\theta=\frac{\dot a_{1}}{a_{1}}+\frac{\dot a_{2}}{a_{2}}+\frac{\dot a_{3}}{a_{3}}\,,\label{eq:def theta}
\end{equation}
\begin{equation}
\sigma^{\,\,\alpha}_{\beta}=\hbox{diag}\,\left[\sigma^{\,\,x}_{x},\sigma^{\,\,y}_{y},\sigma^{\,\,z}_{z},0\right]
=\hbox{diag}\,\left[\Sigma_{1},\Sigma_{2},\Sigma_{3},0\right],\label{eq:def sigma}
\end{equation}
where:
\begin{eqnarray}
\nonumber
\Sigma_{\rm{a}}=\frac{2}{3}\frac{\dot a_{\rm{a}}}{a_{\rm{a}}}-\frac{1}{3}\frac{\dot a_{\rm{b}}}{a_{\rm{b}}}&-&
\frac{1}{3}\frac{\dot a_{\rm{c}}}{a_{\rm{c}}},
 \\
&{}&\!\! \rm{a}\neq \rm{b}\neq \rm{c}\,\left(\rm{a},\rm{b},\rm{c}=1,2,3\right).
\label{eq: componentes}
\end{eqnarray}
The geometric interpretation of these parameters is
straightforward:  $\theta$
represents the isotropic rate of change of the 3-volume of
a fluid element, while
$\sigma^{\,\,\alpha}_{\beta}$ describes its rate of local
deformation along different spatial directions given
by its eigenvectors.
Since the shear tensor is traceless:
$\sigma^{\,\,\alpha}_{\alpha}=0$,
it is always possible to eliminate any one of the three
quantities $\left(\Sigma_{1},\Sigma_{2},\Sigma_{3}\right)$ in
terms of the other two. We choose to eliminate $\Sigma_{1}$
as a function of $\left(\Sigma_{2},\Sigma_{3}\right)$.

Finally, the substitution of (\ref{eq:def theta}) and (\ref{eq:def sigma}) into the Einstein equations, allow us to obtain a first order system of differential equations.

\section{Particles}\label{particles}

At temperature values in the range $100 \,\hbox{MeV}>T>m_e$ neutrons and protons are free in chemical equilibrium.  The equilibrium is possible because of reactions transforming neutrons into protons  and vice versa \cite{Rich}:
\begin{equation} \label{NPreaction}
\nu_e n \leftrightarrow e^- p \, , \qquad \overline{\nu}_e p \leftrightarrow e^+ n \,,
\end{equation}
which implies the following variation of their numbers:
\begin{equation}\label{NPratio}
\frac{n_n}{n_p}=e^{-\frac{\Delta M}{T}}, \qquad \Delta M = m_n - m_p\, .
\end{equation}
%
%
%
On the other hand, for temperature values less than $m_e$ the neutrons decay freely. Hence the densities of neutrons and protons satisfy
%
\begin{equation} \label{NDecayFreel}
\dot{n}_n = - \theta n_n - \Gamma n_n \,.
\end{equation}
 \begin{equation} \label{NNp}
\dot{n}_p = - \theta n_p + \Gamma n_n \,,
\end{equation}
%
where  $\Gamma=1/\tau_n$  is the decay rate ($\tau_n \approx 881.5\, \hbox{s} $).  Also, from the charge neutrality we have $n_{e^-}=n_p$.

Finally, since we assume during the whole evolution a  baryons/photon ratio  $\eta = 5\times 10^{-5}$, we have
\begin{equation}
\frac{n_b}{n_\gamma}=\frac{n_n+n_p}{n_\gamma}=\eta,
\end{equation}
where $n_n$, $n_p$ and $n_\gamma$ are respectively the numbers density of neutrons, protons and photons. In this way we can roughly estimate the neutron and proton concentrations, a necessary task to obtain their rest energy. Since we are not interested in calculating element abundances, this rough approximated result  is sufficient for our purposes. To improve these calculations the magnetic field should be included in the analysis \cite{Dario_review}, but this is beyond the scope of the present paper.
\subsection{Nucleosynthesis constraints on light elements} \label{CBBN}
Since standard calculations of cosmological nucleosynthesis assume an FLRW universe with a radiation equation of state,  it is worthwhile
commenting on the effects of considering an anisotropic universe model on the primordial production of $^4 \hbox{He}$. 
As discussed in \cite{DSysCosm}, the time dependence of the radiation density is very important in determining the helium abundance. 
This yields the following bound for the shear eigenvalues in a Bianchi I model during the nucleosynthesis:
$\hbox{Y}_{\hbox{p}}<0.26$  requires $\left(\sigma/H\right)<0.2$ . 
In the previous equations  $\hbox{Y}_{\hbox{p}}$ and $\hbox{H}$ denote the primordial mass fraction of $^4 \hbox{He}$ 
and expansion scalar, respectively, and $\sigma^2=\left(\Sigma_1^{\,2}+\Sigma_2^{\,2}+\Sigma_3^{\,2}\right)/2$. However, for the above considered 
models with magnetic fields such that $\left< B_0 \right>\lesssim10^{-6} \hbox{G}$, we have that $\left(\sigma/H\right)\ll1$ remains valid throughout the nucleosynthesis process (notice that we assumed the anisotropy of the fluid to be caused only by the magnetic field). 


\end{document}